\documentstyle[12pt,aasms4,psfig]{article}
\def\apj{{\rm Astrophys.\ J.}}

\def\aanda{{\rm Astron.\ Astrophys.}}
\def\aandar{{\rm Astron.\ Astrophys. Rev.}}
\font\sevenbf=cmbx7 
\def\note#1]{{\sevenbf #1 ------]}} 
\def\eg{{\it e.g.}}

\def\etal{{\it et al.}}

\def\kb {k_{\rm B}}
\def\kbt {k_{\rm B}T}

\def\be { \begin{equation} }
\def\ee { \end{equation} }
\def\pd {\partial}

\def\arcsec{{\hbox{\rlap{\rlap{\tt\char"0D}\hbox{\thinspace\tt\char"0D}}
\kern-4.8pt\raise1pt\hbox{$\mit\mathchar"017F$}}}}
\def\arcmin{{\hbox{\rlap{\rlap{\tt\char"0D}}
\kern-5.5pt\raise1pt\hbox{$\mit\mathchar"017F$}}}}
\def\FGH77{Fontaine, Graboske \& Van Horn, 1977}

\def\B91{Baturin, 1991}
\def\EOS92{Vorontsov, Baturin \& Pamyatnykh, 1992}
\def\VB&P92{Vorontsov, S.V., Baturin, V.A. \& Pamyatnykh, A.A., 1992}

\def\bdash{{\kern 0.2em}}

\lefthead{Nayfonov \etal}
\righthead{The MHD equation of state with}

\begin{document}

\title{The MHD equation of state with post-Holtsmark 
microfield distributions}

\author{Alan Nayfonov\altaffilmark{1,2}, 
Werner D\"appen\altaffilmark{1,3,4}, 
David G. Hummer\altaffilmark{5} \& Dimitri Mihalas\altaffilmark{6}} 

\altaffiltext{1}{Department of Physics and Astronomy, University of
Southern California, Los Angeles, CA 90089-1342, U.S.A.}
\altaffiltext{2}{IGPP, Lawrence Livermore National Laboratory,
Livermore, CA 94550, U.S.A.} 
\altaffiltext{3}{Theoretical Astrophysics
Center, Institute for Physics and Astronomy, Aarhus Uni\-versity, 8000
Aarhus C, Denmark} 
\altaffiltext{4}{Institute of Astronomy, Madingley Road, Cambridge CB3
0HA, UK}
\altaffiltext{5}{JILA, University of Colorado and NIST, Boulder CO 80309-0440}
\altaffiltext{6}{Los Alamos National Laboratory, Los Alamos, NM 87545}

\begin{abstract}
The Mihalas-Hummer-D\"appen (MHD) equation of state
is a part of the Opacity Project (OP), where it mainly
provides ionization equilibria and level populations of
a large number of astrophysically relevant species.
Its basic
concept is the idea of perturbed
atomic and ionic states. At high densities, when many-body
effects become dominant, the concept of perturbed atoms loses its sense.
For that reason, the MHD equation of state was originally restricted to
the plasma of stellar envelopes, that is, to relatively moderate
densities, which should not exceed $\rho < 10^{-2}$~g~cm${}^{-3}$.
However, helioseismological analysis has demonstrated that
this restriction is much
too conservative. The principal feature of the 
original \cite{hm88} (1988) paper is
an expression for the
destruction probability of a bound state 
(ground state or excited) of a species
(atomic or ionic), linked to the
mean electric microfield of the plasma.~\cite{hm88} (1988) assumed,
for convenience,
a simplified form of the
Holtsmark microfield for randomly distributed ions.
An improved MHD
equation of state (Q-MHD) is introduced. 
It is based on a more realistic
microfield distribution (\cite{ho66-68} 1966, 1968)
that includes plasma correlations.
Comparison with an
alternative post-Holtsmark formalism (APEX)
is made, and good agreement is shown.
There is a clear signature of the choice of the
microfield distribution in
the adiabatic index $\gamma_1$, which makes it
accessible to present-day
helioseismological analysis.
But since these thermodynamic effects of
the microfield distribution
are quite small, it also follows that
the approximations chosen in the original MHD
equation of state were reasonable.
A particular feature of the original MHD papers was an explicit
list of the adopted free
energy and its first- and second-order analytical derivatives.
The 
corresponding Q-MHD quantities are given in the Appendix.
\end{abstract}


\section{Introduction}

The so-called MHD equation of state 
(\cite{hm88} 1988;~\cite{mdh88};~\cite{dmhm88};~\cite{dam87})
was developed as part
of the international ``Opacity Project'' 
(OP; see~\cite{sea87-95};~\cite{ber97}).
Its main
purpose was to calculate the ionization degrees of all astrophysically
relevant chemical elements in order to provide a crucial ingredient of the
calculation of the radiative opacity of stellar interiors. The basic
concept of the MHD equation of state was built on the idea of perturbed
atomic and ionic states. At high densities, when many-body
effects become dominant, the concept of perturbed atoms loses its sense.
For that reason, the MHD equation of state was originally restricted to
the plasma of stellar {\it envelopes}, that is, to relatively moderate
densities, which should not exceed $\rho < 10^{-2}$~g~cm${}^{-3}$.

However, the MHD calculation of ionization equilibria was not only the
necessary part of an opacity calculation. The same analytical and
computational effort also allowed the computation of thermodynamic
quantities to a high degree of accuracy and reliability.  It turned out
that for the purpose of thermodynamic calculations, the aforementioned
density domain was much too conservative. For 
instance,~\cite{cdl88} (1988)
applied the MHD equation of
state to models of the entire Sun in order to predict solar
oscillation frequencies. MHD remained a reliable tool down to the solar
center, where density is about 150~g~cm${}^{-3}$. So in spite of the
original design of the MHD equation of state,
the
associated thermodynamic expressions have a broader domain of
applicability, extending in particular to stellar {\it cores}. The reason
is that in the deeper interior, the plasma becomes virtually fully
ionized. Therefore, in practice, it does not matter that the condition
for the legitimacy of the perturbation mechanism for bound species
(\cite{hm88} 1988) is not fulfilled, essentially
because there are no bound species of the chemical elements that 
can be relevant
for the equation of state (however, for {\it opacity},
bound states of less abundant elements are relevant).
Without bound species, the MHD
equation of state falls back to an ideal-gas equation, enriched with
Coulomb pressure and electron degeneracy. Coulomb pressure is only
included to lowest order (the Debye-H\"uckel approximation, see,
\eg, \cite{ekk76}), yet it is now known that, somewhat fortuitously, the
lack of higher-order Coulomb term has no significant consequence (for
more details see~\cite{cd96}). Specifically, it was shown 
by~\cite{cb97} (1997) that the MHD equation of state can be
used for low-mass stars, at least down to 0.4~M${}_{\odot}$ (see
also~\cite{td99}). 

%%%%%%%%%%%% do not justify para (%-sign!!)  
%%%%%%%%%%%% do not justify para (%-sign!!)  
Helioseismology has so far been putting the toughest observational
constraints on the equation of state (for a recent
review, see~\cite{cddg99}). Given the importance of
helioseismology as an experiment, we specifically
review some of the most recent
results in Section~3. It has emerged that 
the MHD equation of state
clearly fares well, especially in comparisons with simpler formalisms,
for instance the popular~\cite{eff73} (1973) 
equation of state (hereinafter EFF). However, there are
still discrepancies between MHD and the inference from 
observations (\cite{cd96}). 
Some of
these discrepancies were successfully
removed by the OPAL equation of state (which is
itself part of a major opacity 
project;~\cite{ro86};~\cite{ir95} 1995;~\cite{rsi96}
and references therein). However,
recent results by \cite{bdn99} (1999) have indicated that in
the outermost
layers of the Sun ($r/R_{\odot} > 0.97$), 
the MHD equation of state
appears to be a better match to the data than 
OPAL (see Section~3). 

The continual 
involvement of the MHD equation of state in current developments
has led us to revisit the principal issues. The
original \cite{hm88} (1988) paper derived an expression for the 
destruction probability of a bound state (ground state or excited) of a species
(atomic or ionic). The probability was expressed as a function of the
mean electric microfield. For convenience, \cite{hm88} (1988) assumed a
Holtsmark microfield for randomly distributed ions. Furthermore, to
reduce the computational effort, they adopted a simplified approximate form
of the Holtsmark function.

It turned out that these simplifications
have been the source of inadequacies of
the MHD equation of state. In a real plasma, correlations occur, that
is, the Coulomb interaction modifies the ion distribution. The result
is that the microfield distribution peaks at lower values of the field
strength than given by the Holtsmark distribution. Therefore, the
probability with which a state for a specific atom or ion ceases to
exist is reduced relative to the Holtsmark result, and thus finally, the
occupation probabilities is higher than in the Holtsmark approximation. 
These discrepancies were demonstrated by \cite{ir95} (1995).
Therefore, the poor quality of the Holtsmark approximation is one of the
reasons for the discrepancy between OP and OPAL opacities at higher
densities.

One might think that these differences would have only a small bearing
on the equation of state (\cite{ri98} 1998). 
However, recent studies about
the influence of details in the hydrogen partition function on
thermodynamic quantities 
(\cite{nd98} 1998;~\cite{bdn99} 1999) 
have revealed that
the difference between various approximations of the microfield
distribution is well within range of observational helioseismology
(see Section~3).
In view of these more stringent demands on the equation of state,
we have improved the original MHD equation of state by including a
post-Holtsmark microfield distribution to account for particle
correlation. This microfield distribution function was derived 
by~\cite{ho66-68} (1966, 1968). 
Here, we present the resulting improved MHD
equation of state. We name it ``Q-MHD'' after~\cite{h86}'s (1986)
nomenclature (in which the microfield
distribution was called~P and its integral~Q).
Comparison with an
alternative post-Holtsmark formalism (APEX; see \cite{idwlmgh85} and
references therein) is made, and good agreement is shown. Since one of
the main features of the original MHD papers was an
explicit list of the free
energy and its first- and second-order analytical derivatives, we list
all the corresponding Q-MHD quantities in the Appendix. 

\section{The MHD Equation of State}

The majority of the realistic equations of state that have appeared in
the last 30 years are based on the so-called ``free-energy minimization
method''.  Before that, a common method was to use one expression
({\it e.g.}, a modified Saha equation) 
for the ionization equilibria and another
(not necessarily consistent) expression for the
thermodynamic quantities.  Such a procedure can lead to violations
of Maxwell
relations. The prevention is of course the use of a single thermodynamic
potential, {\it e.g.}, the free energy. Then, thermodynamic 
consistency is
built in. However, since such a method requires
considerable computing power, especially for multi-component plasmas, 
it could only become feasible since about 1960
(\cite{ha59};~\cite{hrt60}). 

The free-energy minimization method uses statistical-mechanical
models (for example,
a partially degenerate electron gas, Debye-H\"uckel theory for ionic
species, hard-sphere atoms to simulate pressure ionization via
configurational terms, quantum mechanical models of atoms in perturbed
fields, {\it etc.}). It is a modular approach, that is, these models
become the building blocks of a macroscopic free energy, which is
expressed as a function of temperature $T$, volume $V$, and the particle
numbers $N_1, \ldots, N_m$ of the $m$ components of the plasma.  This
model free energy is then minimized, 
subject to the stoichiometric constraints. 
The solution of this minimization
problem then gives both the equilibrium
concentrations and, if inserted in the free energy and its derivatives,
the equation of state and the thermodynamic quantities. Obviously, this
procedure automatically guarantees thermodynamic consistency.
For obvious reason, this approach is called the ``chemical picture''. 
Perturbed atoms must be introduced on a more or less {\it ad-hoc} basis
to avoid the familiar divergence of internal partition functions (see,
{\it e.g.}, \cite{ekk76}).

\subsection{Internal Partition Functions}

The simplest way to modify atomic or ionic bound states to account for
effects of the surroundings is by truncating the internal partition
function at some maximum level that can be a function of temperature
and density. Such a modification of bound states runs into the
well-known technical problem that whenever the density passes through a
critical value, for which a given bound state disappears into the
continuum, the partition function changes discontinuously by the amount
of the statistical weight $g_i$ of the state. This is clearly unphysical
and would lead to discontinuities and singularities in the free energy
and its derivatives.

One way of avoiding the problem of discontinuous jumps in the partition
function is to assign ``weights'' or ``occupation probabilities'' to all
bound states of all species. The internal partition function
then becomes

\be
Z_{jk}^{\rm int} = \sum_{i}{w_{ijk}}~{g_{ijk}}
{\exp\left[-E_{ijk}/(\kbt)\right]} \ .
\label{E.1}
\ee
Here, $\kb$ is the Boltzmann
constant, $T$ temperature,
$w_{ijk}$ the probability that the state $i$ of ion $j$ of species
$k$ still exists despite the plasma environment, and 
$g_{ijk}~e^{-E_{ijk}/(\kbt)}$ indicates the probability 
that this state is actually occupied in the system.
Such an occupation probability formalism has several advantages:

\noindent
$\bullet$ The $w_{ijk}$ decrease continuously and monotonically as the
strength of the relevant interaction increases. 

\noindent
$\bullet$ States now fade out continuously with decreasing $w_{ijk}$ and thus
assure continuity not only of the internal partition function but
also of all material properties (pressure, internal energy, etc.)

\noindent  
$\bullet$ The probabilistic interpretation of $w_{ijk}$ allows us to combine
occupation probabilities from statistically independent interactions. It
is thus straightforward to allow for the simultaneous action of
different mechanisms, accounting for several different species of
perturbers by any one mechanism. Hence the method provides a scheme for
treating partially ionized plasmas, and the limits of completely neutral
or completely ionized gas are smoothly attained. 

\noindent
$\bullet$ The $w_{ijk}$ can be made analytically
differentiable. In this way, MHD realized a reliable second-order 
numerical scheme in the free-energy minimization. 

\noindent
$\bullet$ Finally, the $w_{ijk}$ formalism can be related naturally 
to line-broadening theory, which is important both for the interpretation 
of laboratory spectra and for opacity calculations.

\subsection{Statistical-Mechanical Consistency}

Although the $w_{ijk}$ can only be calculated {\it a priori} from some 
complementary interaction model, it must be stressed that one cannot 
introduce occupational probabilities into the internal partition function
completely arbitrarily.

\cite{fer24} (1924) showed that one can {\it derive} the
$w_{ijk}$ from a free energy that depends explicitly on the individual
occupation numbers $N_{ijk}$; conversely it follows that the use of
some heuristic $w_{ijk}$ in the internal partition function {\it
implies} the existence of an equivalent nonideal term in the free energy
$F$ (see~\cite{fow36}). 

To illustrate that particular care must be taken when using $w_{ijk}$,
consider a single-species perfect gas (that is, 
no dissociation or ionization),
with a total of $n$ particles, of which $n_i$ are in state
$i$ (thus $n = \sum n_i = {\rm const}$).

If $f$ is {\it defined} by

\be
w_i \equiv e^{ -(\pd f/ \pd n_i)/(\kbt)}\ ,
\label{E.2}
\ee
then, to get a consistent free energy, the internal partition function
$Z^{\rm int}$ [Eq.~(\ref{E.1})] 
must include the weights $w_i$ of Eq.~(\ref{E.2}).  
We denote by $Z^{*}$ such a particular internal partition function. 
For statistical-mechanical consistency, the free energy must contain a
corresponding external part, which is
the last term in the following equation for the free energy of the gas
(\cite{fow36})

\be
F = - \kbt n
\left[{\log\left({ {V\over n} {1\over{\lambda^3}} }\right) + 1}\right] -
\kbt n \log Z^{*} + 
\left[{ f - \sum_{i} n_i (\pd f/ \pd n_i) }\right] \ ,
\label{E.3}
\ee
where $V$ is the volume occupied by the gas, and
$\lambda$ is the de Broglie wavelength of the particles of
the species considered here.

The last term in this equation has in fact been often ignored without
justification. However,
If the interaction term happens to be {\it linear} in the $n_i$,
then the last bracket in Eq.~(\ref{E.3}) vanishes, and to get a consistent
formulation it is sufficient merely to use the appropriately
modified internal
partition function $Z^{*}$ in the standard expression for $F$. 
Obviously, the bracket will not, in general, vanish for interaction
terms that contain
nonlinear combinations of the $n_i$. However, in such cases it might be
possible to find an astute simplified linear version of the interaction
model which would reduce the problem to the previous case.

\subsection{MHD Occupational Probabilities}

\subsubsection{Perturbations by Neutral Species}

For neutral perturbing species, \cite{hm88} (1988)  started out
from a widely studied hard-sphere model (\cite{fow36}) with each state
in principle having its own diameter.  However, the simple binary
interaction model is computationally prohibitive because it accounts for
perturbations from all ions of all chemical species in {\it all}
possible excited states. That implies thousands of the
{\it individual} occupation numbers $N_{ijk}$ as independent variables
in the free energy minimization.  In addition, in this case the function $f$ is
nonlinear.

As an obvious first approximation, MHD considered the 
{\it low-excitation limit} (see \cite{hm88} 1988) in which it is assumed that
essentially all perturbers encountered by an atom in an exited state
reside in the ground state. 

\be {(w_{ijk})}_{\rm neutral} = \exp \left[ {
-(4\pi /3V) \sum_{j',k'} N_{j'k'} {(r_{ijk} + r_{1j'k'})}^3} \right] \ . 
\label{E.4}
\ee
Here, the $r_{ijk}$ are the radii of state $i$ of ion $j$ of species $k$.
Thus, the problem is reduced to {\it total} occupation numbers $N_{jk} =
\sum_{i} N_{ijk}$
of all ionic species. And by eliminating the explicit appearance of
$N_{ijk}$ the interaction $f$ becomes linear in individual occupation
numbers,

\be
f_{\rm neutral} = \kbt \left({4\pi \over{3V}}\right) \sum_{i,j,k} N_{ijk}
\sum_{j',k'} N_{j'k'}
{(r_{ijk} + r_{1j'k'})}^3 \ .
\label{E.5}
\ee
Therefore the term of the form

$$\left[{ f - \sum_{i} n_i (\pd f/ \pd n_i) }\right] \ ,$$

\noindent
appearing in the multispecies generalization of Eq.~(\ref{E.3}) vanishes
identically, and the interaction only appears in the factors
$w_{ijk}$,
that is, in $Z^{*}$.

\subsubsection{Perturbations by Charged Species}
                                                                
In the case of charged particles, instead of trying to figure out $f$
itself, MHD defines $w_i$ directly, arguing that the presence of a
plasma microfield destroys high-lying states by means of a series of 
Stark level mixing with higher lying states leading to the continuum.

The basic idea is that for each bound state of every unperturbed
ion (labeled by the indices $i,j,k$), 
there is a critical value of the electric field $F_{ijk}$ such that
the state in question cannot exist if the field exceeds the critical
value. Then probability that a given state {\it does} exist is simply
the probability that the field strength is less than $F_{ijk}$, i.e.,

\be {(w_{ijk})}_{\rm charged} = {\int}_{0}^{F_{ijk}} P(F) dF \ ,
\label{E.6}
\ee 
where $P(F)$ is the microfield distribution function (see Section~4). 
The choice of an appropriate plasma microfield $P(F)$ is not
straightforward. \cite{hm88} (1988) have made the
following choice (see section~4.2) 
of the resulting $w_{ijk}$, based on numerical comparisons with existing
atomic physics calculations

\be {(w_{ijk})}_{\rm charged}
= \exp \left\{ { -\left( {4\pi \over{3V}}\right)16 {\left[ {
{(Z_{jk}+1)}^{1/2} e^2\over{ {K}_{ijk}^{1/2} \chi_{ijk}} }\right]}^3
\sum_{j',k'} N_{j'k'} {Z}_{j'k'}^{3/2} }\right\} \ ,
\label{E.7}
\ee 
where $Z_{jk}$ denotes the charge of ion $j$ of chemical
species $k$ (thus, zero for neutral particles) and the sum runs as
before over all
levels~$i$ of ions~$j$ of species~$k$. $K_{ijk}$ is a quantum
correction factor of those levels (see Eq.~4.70 of \cite{hm88} 1988).
Note that the interaction due to $w_{\rm charged}$ is automatically 
{\it linear} in the abundances
$N_{ijk}$ because this model of interaction does not depend on the
internal excitation states of the perturbers. Therefore, there is
again
no term of the form of the last term in
Eq.~(\ref{E.3}), and the interaction only appears in the factors $w_{ijk}$.
We stress that this
property was of fundamental importance for the numerical feasibility of
the MHD equation of state (\cite{hm88} 1988). 

Assuming statistically independent actions from neutral and charged
perturbers, the joint occupation probability is the product of
${(w_{ijk})}_{\rm neutral}$ and ${(w_{ijk})}_{\rm charged}$.  Formally, the
neutral term could also be retained when extended charged particles
interact with neutrals. However, because a hard-sphere description is
highly implausible when one of the particles is charged, the MHD
equation of state restricts the use of ${(w_{ijk})}_{\rm neutral}$ for
mutually neutral species ($jk$) only. 

\section{Helioseismic Equation-of-State Diagnosis}

\begin{figure}
\figcaption{Difference between squared sound speed from inversion
of oscillation data and that of a standard model based on the MHD (circles) 
and OPAL (triangles) equation of state (Figure provided by S. Basu).
\label{fig-hel1}}
\vspace{2cm}
\psfig{file=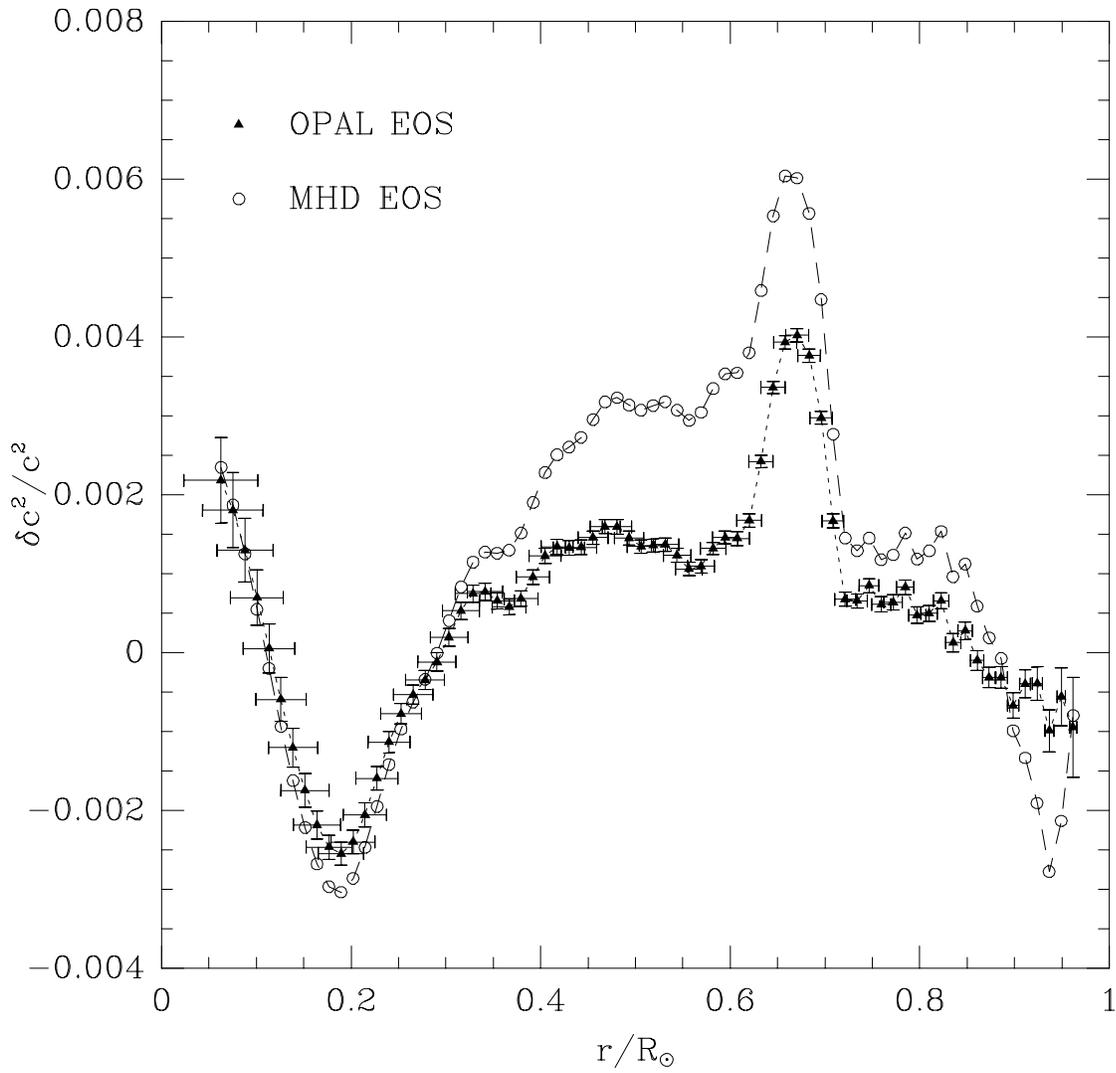}
\end{figure}

In the solar convection zone, helioseismology presents an
opportunity to isolate the question of the equation of state
from opacity and nuclear reaction rates, since the
stratification is essentially adiabatic and thus determined
by thermodynamics (see, \eg, \cite{cd92}). Accurate analysis
of the observations requires use of the full, nonasymptotic
behavior of the oscillations. Figure~\ref{fig-hel1} 
shows a typical
result of a numerical inversion (\cite {bc97}). It shows the
relative difference (in the sense Sun -- model) between the
squared sound speed obtained from inversion of oscillation
data and that of a two standard models. A perfect model
would lie on the zero line. The two models are in all
respects identical except that they use a different equation
of state (MHD and OPAL, respectively, {\it cf.} Section~1).
Simplifying for the present purpose, we can look at results
such as shown in Figure~\ref{fig-hel1} as the {\it data} of
helioseismology, disregarding how they were obtained from
solar oscillation frequencies.

\placefigure{fig-hel1}

The most important result of the earlier helioseismic
equation-of-state analyses (\cite{cd96}) was that it is
essential to include the leading Coulomb correction (the
Debye-H\"uckel term) to ideal-gas thermodynamics. Under
solar conditions, the size of the relative Coulomb pressure
correction is largest in the outer part of the convection
zone (about~--8 percent) and it has another local maximum in
the core (about~--1 percent). Figure~\ref{fig-hel1}, 
however, hides the
fact that the Coulomb correction is the most important one,
since both MHD and OPAL already contain it. Note that in
Figure~\ref{fig-hel1} the most significant 
information about the equation
of state regards the convection zone ($r > 0.71 R_{\odot}$),
since beneath it, one cannot disentangle the influence from
the equation of state from other effects. 
Figure~\ref{fig-hel1} contains
the evidence that in the region $0.90 R_{\odot} < r < 0.97
R_{\odot}$, OPAL is a better fit to
reality than MHD.

Two very recent inversions have had further implications for
the equation of state. First, the strong constraints from
helioseismology now force us to include {\it relativistic
effects} of electrons (\cite{ek98}). 
Neither MHD nor OPAL have so far
included relativistic effects, unlike the earlier and
simpler EFF (Section~1) and its more recent sibling SIREFF
(\cite{gs97}). Second, there are indications that for
$r > 0.97 R_ {\odot}$, MHD is favored over OPAL. This is the
result of an apparent helioseismic confirmation of the
thermodynamic effect of the excited states in hydrogen,
treated according to MHD [Eq.~(4),(6)]. The
effect itself was demonstrated in a theoretical study 
by~\cite{nd98} (1998), which revealed interesting features beyond the
Coulomb correction approximation. They are related to
excited states and their treatment in the equation of state.
The MHD equation of state with its specific,
density-dependent occupation probabilities (Section~2.3) 
is causing a characteristic ``wiggle'' in the
thermodynamic quantities, most prominently in 
$\chi_{\rho} = (\partial \ln p /\partial \ln \rho)_{T}$, 
but equally present in the other thermodynamic quantities.

\placefigure{fig-hel2}

For convenience, in Figure~\ref{fig-hel2},
we recall the result for the
helioseismically relevant $\gamma_1$ of a hydrogen-only
plasma. Temperatures and densities are taken from a
solar model. The appropriate density is implied but not
shown (for more details and different
thermodynamic variables and chemical compositions, see
\cite{nd98} 1998 and Sections~6.2, 6.3). Five cases were considered: 
(i) ${\rm MHD}$ [standard MHD occupation probabilities:
Eq.~(4),(6)], 
(ii) ${\rm MHD_{GS}}$ [standard MHD internal
partition function of hydrogen but truncated to the ground
state (GS) term], 
(iii) ${\rm OPAL} $: OPAL tables [version
of November 1996 of \cite{rsi96} (1996)], 
(iv) ${\rm MHD_
{PL}}$ [MHD internal partition function of hydrogen, but
replaced by the Planck-Larkin partition function
$Z_{PL} = \sum_1^{\infty} g_n 
[\exp(-E_n/\kbt) + E_n/\kbt - 1]$ (\cite{ro86})],
(v) ${\rm MHD_ {PL,GS}}$ [${\rm MHD_{PL}}$ truncated to the
ground state term]. The effect of the inclusion of the
excited states in the internal partition function is
manifest in the differences between MHD and ${\rm MHD_{GS}}
$, and between $ {\rm MHD_{PL}}$ and ${\rm MHD_{PL,GS}}$,
respectively. The effect of different occupation
probabilities of ground and excited states shows up in the
difference between MHD and $ {\rm MHD_{PL}}$, and between $
{\rm MHD_{GS}}$ and ${\rm MHD_{PL,GS}}$. It was found that
the presence of excited states is crucial. Also, the wiggle
was demonstrated to be a genuine neutral-hydrogen effect
despite the fact that most hydrogen is already ionized. This
qualitative picture does not change when helium is added
\cite{nd98} (1998).

\begin{figure}
\figcaption{Relative difference
between $\gamma_1$
obtained from an inversion of
helioseismological data and $\gamma_1$ of
the solar models listed in Table~I,
in the sense ``Sun -- model''.
Only the {\it intrinsic} difference in $\gamma_1$ is shown, 
that is, the part of the difference due to the equation of state
(Figure provided by S. Basu).
\label{fig-hel3}}
\vspace{4cm}
\psfig{file=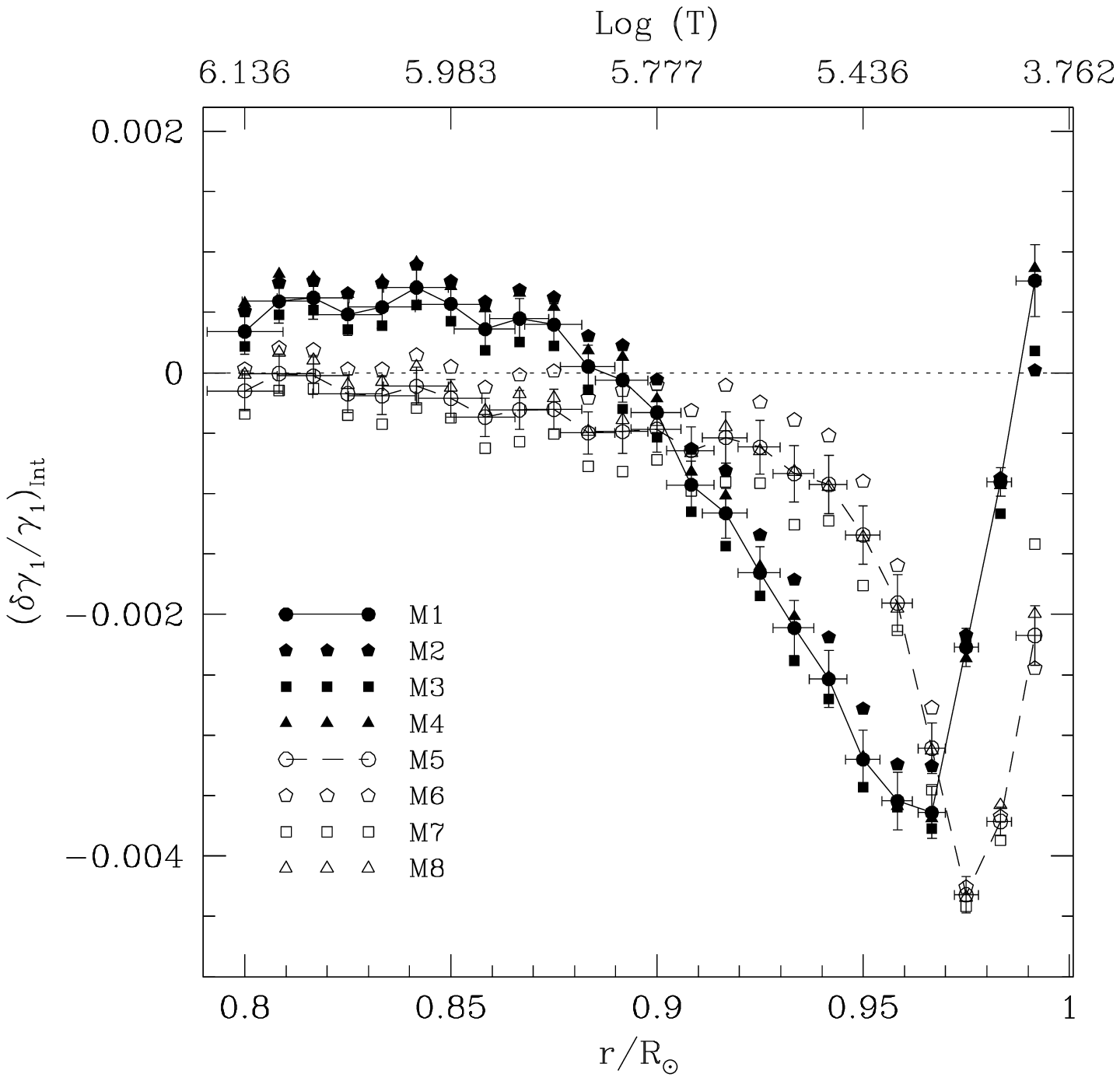}
\end{figure}

It seems that this effect of excited states can be
observed in the Sun. Figure~\ref{fig-hel3} 
shows the result of the inversion 
by \cite{bdn99} (1999), based on the solar
oscillation frequencies obtained from the SOI/MDI instrument
on board the SOHO spacecraft during its first 144 days in
operation (\cite{rh97} 1997). Previous sound-speed and
$\gamma_1$ inversions (\eg, \cite{ek98}) 
were still indirect, that is, they were giving the difference
between solar models and the Sun without separating the
change in structure due to the equation-of-state contribution.
The inversion by \cite{bdn99} (1999)
was done for the
so-called ``intrinsic'' $\gamma_1$ difference between solar
models and the Sun. For convenience, we repeat 
the principal result
in Figure~\ref{fig-hel3}, which 
shows the
difference between the Sun and various calibrated solar
models. The models alternate between two equations of state
(MHD, OPAL), three different values for the solar radius,
and two formalisms for convection [MLT: standard mixing
length theory, CM: \cite{cm91} (1991) formalism]. The models are
specified in Table~1, where in addition the calibrated
values of the surface helium abundance $Y_s$ and the depth
of the convection zone $r_{\rm cz}$ are listed. All models
assume gravitational settling of helium and heavy
elements, an effect that is now part of the so-called standard solar
model (\cite{cd96}).

\placefigure{fig-hel3}

In contrast to earlier results for the intrinsic $\gamma_1$
difference (\cite{bc97}), which had uniform resolution
throughout the Sun, the new study focuses on the 20\%
uppermost layers. It appears that in the top layers, the MHD
models give a more accurate description of the Sun than the
OPAL models. Since the difference in $\gamma_1$ between MHD
and OPAL is the wiggle of Figure~\ref{fig-hel2}, 
the observed preference of
the MHD model in the upper region could indicate a
validation of an MHD-like treatment of the exited states.
Fig.~\ref{fig-hel3} confirms the aforementioned earlier findings that
below the wiggle region, OPAL fares better than MHD. In that
region, the Planck-Larkin partition function of OPAL 
appears to be the better choice.

The results in favor of MHD in the upper part of the Sun
($r > 0.97R_{\odot}$) could be due to the different
implementations of many-body interactions in the two
formalisms. Since density decreases in the upper part, OPAL
by its nature of a systematic expansion, inevitably becomes
itself more accurate; but MHD might, by its heuristic
approach, have incorporated even finer,
higher-order effects.

Let us add a word of caution, though. It could appear
tempting to produce a ``combined'' solar equation of state,
with MHD for the top part and OPAL for the lower part.
However, such a hybrid solution is fraught with danger. For
instance, it is known that patching together equations of
state can introduce spurious effects (\cite{da93}). It seems
that the right way is to improve MHD and OPAL in parallel
and independently, guided by the progress of
helioseismology. The aim of the present study is to improve MHD.

\section{Microfield Distributions}

One of the mechanisms that affects the bound states of a {\it radiator}
(atom or complex ion immersed in a plasma) is the electric field
arising from the charged particles in its environment.  

To be able to calculate the internal partition function it, therefore,
is necessary to estimate $E_{nlm}$ and to calculate a microfield
distribution. The actual electric field at a radiator in plasma is
fluctuating in time. From the time-scale point of view, it is convenient
to split the electric field into two components:  a high-frequency part
(with respect to the time scale of particle collisions) and a
low-frequency component which varies on a time scale much longer than
the orbital period of the bound state considered. Because of their high
velocities, free electrons are considered as perturbers in the
high-frequency part of the microfield distribution only. The
distribution of the low-frequency component is calculated by considering
a gas of ion perturbers. As was shown in (\cite {hm88} 1988), the
low-frequency component dominates microfield effects at least for the
conditions of stellar envelopes.  In the following we discuss only this
component. 

\subsection{Holtsmark Microfield Distribution}

\cite{u48} (1948) suggested a model for a microfield distribution
that allowed for a simple analytical formulation. It contained the {\it
nearest neighbor} (NN) approximation and, therefore, was better suited
to study the limit of strong fields. The long range of Coulomb
interactions in plasmas, however, renders the nearest neighbor
approximation inadequate.  The next logical step in the development of
the microfield distributions should involve the interaction of the
radiator with {\it all} ions in plasma. For a case of pure hydrogen
plasma this problem was worked out by \cite{h19} (1919). Two
serious shortcomings of the Holtsmark distribution are:

\noindent
$\bullet$ its limitation to neutral radiators and

\noindent 
$\bullet$ the absence of correlations between the charged plasma
perturbers.

The original MHD formalism (\cite{hm88} 1988) modified the Holtsmark
distribution by making plausible but non-rigorous correction for the
effects of ions other than protons. It considers hydrogenic radiators in
a plasma for which the microfield perturbations by a variety of ionic
species of charge $Z_p$, $p = 1,2,$ are dominant. The resultant
occupation probability of an electron in level $i$ = ($n,l,m$) of a
hydrogenic potential of charge $Z_a$ is given by (see Eq.~4.68a of 
\cite{hm88} 1988) 

\begin{equation}
w_{i}^{c} = Q\left[ { {K_i \chi_i^2 }\over{4 Z_a a_0^3} } 
{\left({4\pi n_e\over 3}\right)}^{-2/3}
{\left(n_{ion}\over n_e\right)}^{1/3}\right] \ ,
\label{E.8}
\end{equation}
where

\begin{equation}
Q(x) = \int_0^x P_H(\beta) d\beta \ ,
\label{E.9}                                     
\end{equation}
$K_i$ is the quantal Stark-ionization correction factor, $\chi_i$ is the
ionization potential of the level $i$, $a_0$ is the Bohr radius, $n_e$
is the electron density, $n_{ion}$ is the density of all ions, and
$P_H(\beta)$ is the Holtsmark distribution function

\begin{equation}
P_H(\beta) = \left( 2\beta\over\pi\right){\int_0^\infty} dy 
\exp(-y^{3/2})y \sin\beta y.
\label{E.10}
\end{equation}

\subsection{MHD Microfield Distribution}

In the form of Eq.~(\ref{E.8}) the calculations of plasmas with dozens
of ions and hundreds of bound states are too cumbersome, especially if
one wants to code all the derivatives in the analytical form. Instead,
MHD relied on numerical experiments in search for analytical fits that
would mimic the Holtsmark distribution function. It turned out that a
good starting point was the~\cite{u48} (1948) 
occupation probability $w_i^{c}$,
given by

\begin{equation}
w_i^{c} = \exp \left( -{32\pi\over {3V}}{Z_a^{3/2} a_0^3\over{K_i^{3/2} 
\chi_i^3}}\sum_p N_p Z_p^{3/2} \right).
\label{E.11}
\end{equation}
It might be worth noting here that because $\chi_{i} \propto 1/n^{2}$,
even putting the arbitrary factor of 2 into exponent of Eq. (11) changes the
cutoff quantum number by only a factor of $2^{1/6} = 1.12$, which is
typically quite negligible.

This simple expression is indeed a good fit to the Holtsmark
distribution, if an additional more-or-less {\it ad hoc} factor~2 is put
in the argument of Eq.(\ref{E.8}). Tests showed (\cite{hm88} 1988)
that for strongly bound states with $w_i^{c} \geq 0.1$ this trick
actually leads to a good fit. However, for $w_i^{c} \leq 0.1$ this fit
decreases much more sharply than does Eq.(\ref{E.8}); for
many applications this is not a matter of serious concern inasmuch as
the basic physical effect - that such levels are for all practical
purposes destroyed - is achieved.

The adopted form of $w_i^c$ in MHD equation of state reflects the
interactions of the radiator with all ions in the plasma, but the
approximation does not take into account the correlations of the charged
perturbers. This effect is already a problem because of the choice 
of the Holtsmark
distribution function, not only one of the replacement of the Holtsmark
distribution by a modified Uns\"old expression (\cite{ir95} 1995).

\subsection{Microfield Distributions with Correlation Effects}

\subsubsection{APEX Microfield Distribution}

One of the alternative methods to calculate the microfield distribution
for plasmas of general nature, which include particle correlations, is
the APEX (``adjustable-parameter exponential'') approximation ( see
\cite{idwlmgh85} and references therein) based on a so-called
``independent quasiparticle model''.

Originally developed as a phenomenological method, APEX was shown to be
in agreement with a rigorous theoretical procedure (\cite{dbi85})
related to a Baranger-Mozer series type of analysis. APEX has been shown
to agree quite well with computer Monte-Carlo simulations for both high-
and low-frequency component distributions.  Especially well-suited for
high-Z plasmas (\cite{ilmg83}), 
it treats the low-frequency component by considering a
gas of ions interacting through electron screened potentials, which is a
way to include static contributions from both ions and electrons into
account.  The success of APEX can be attributed to a fact that for small fields
the contributions from many ions are important and these are well
characterized by the second moment of the microfield distribution
that is exactly included in APEX. 

The correlation effects are incorporated through the radial distribution
functions of the ionic perturbers calculated in the hyper-netted-chain
approximation generalized to multi-component plasmas (\cite{r80}). 

\subsubsection{Q-fit Microfield Distribution}

An alternative improvement over the Holtsmark distribution function,
which accounted for particle correlations yet remained sufficiently simple 
so that it could be expressed by analytical fits suited to an         
equation-of-state program, was introduced by one of the authors (D.
Hummer) in 1990. 
The numerical fits were based on work by~\cite{ho66-68} (1966, 1968).
Although until the present work these fits have never been 
included in
equation of state 
calculations, they were successfully used in an
non-LTE investigation of the stellar flux near the
series limits in hot stars
(\cite{hub94}).

The~\cite{ho66-68} (1966, 1968)
microfield distribution was derived under the following assumptions:

\noindent
$\bullet$ the perturbing ions and electrons are in equilibrium with the
same kinetic temperature,

\noindent
$\bullet$ while multiply charged radiators were allowed, all of the
perturbing ions were singly charged. 

While these assumptions are obviously too stringent for studies of
general plasmas and even laser-generated plasmas (see \cite{th77}),
they should be quite reasonable for the plasma of the interior of more
or less normal stars having typical astrophysical compositions. In
this case, one can show that  the two assumptions of the Hooper
microfield  are indeed satisfied. First, the {\it interior} of normal
stars is in thermodynamic equilibrium. Electrons and ions have
therefore the same temperature. Note that stellar {\it atmospheres}
are of course not in  thermal equilibrium, but any stellar equation of
state of the type discussed in this article is inadequate for them,
and special nonlocal formalisms are needed. Second, for not-too-evolved
stars, the typical chemical composition of the universe
prevails, which has about 90\%
hydrogen by number. For such plasmas, the vast majority of
perturbing ions are protons, {\it i.e.}, singly charged.

When plasma correlation effects are important, the microfield distribution
function depends on two additional parameter, the radiator charge $Z_r$ and
the correlation parameter

\begin{equation}
a = {\left({4\pi\over 3} n_e r_D^3\right)}^{-1/3} 
= \ { 0.09 n_e^{1/6}\over {T^{1/2}} } \ ,
\label{E.12}
\end{equation}
where $n_e$ is in ${\rm cm}^{-3}$ and $T$ is in $K$.

\noindent
The physical meaning of the parameter $a$ is clear from Eq.~(\ref{E.12}),
namely $a = \eta^{-1/3}$, where $\eta$ is the number of ions in a Debye
sphere of radius $r_D$

\begin{equation}
r_D = \sqrt{\kbt \over {4\pi e^2 n_e} }.
\label{E.13}
\end{equation}

In the low-frequency approximation only electrons contribute to
shielding [Eq.~(\ref{E.13})] 
and the overall charge neutrality of plasma was
used to obtain the last form of Eq.~(\ref{E.12}).

Another way to look at $a$ is to realize that it can be expressed in
terms of the electron coupling parameter $\Gamma_{\rm e}$: 

\begin{equation}
a^2 =3 \left({ e^2\over{\kbt a_e} }\right) = 3 \Gamma_{\rm e} \ ,
\label{E.14}
\end{equation}
where $a_e$ is the electron sphere radius defined by
${4\pi\over 3} n_e a_e^3 = 1$.

The appropriate microfield distribution function $W(\beta;Z_r,a)$ has been
derived by~\cite{ho66-68} (1966, 1968). 
Using a substantially modified version of Hooper's
code, Thomas Sch\"{o}ning of the University of Munich Observatory 
and Hummer computed two-dimensional fits in $\beta$
and $a$ to $W(\beta;Z_r,a)$ for $Z_r =0, \ldots, 5$ and $a \le 0.8$. From
these fits Hummer evaluated the function defined by Eq.~(\ref{E.9}) 

\be
Q(\beta; Z_r, a) \equiv \int_{0}^{\beta} W(\beta';Z_r,a) d\beta'
\label{E.15}
\ee
for the given ranges of the parameters $Z_r$ and $a$.
It was possible to get a reasonably good analytical fit to obtained data
by using just two parameters: radiative charge $Z_r$ and a correlation
parameter $a$.

The adopted form of the fit $Q(\beta,Z_r,a)$ is 
\begin{equation}
Q(\beta,Z_r,a) = { f(\beta,Z_r,a)\over{1 + f(\beta,Z_r,a)} },
\label{E.16}
\end{equation}
where 
\begin{equation}
f(\beta,Z_r,a) = {C_1(Z_r,a) \beta^3 \over{ 1 + C_2(Z_r,a) \beta^{3/2}} }.
\label{E.17}
\end{equation}

The coefficients $C_1$ and $C_2$ depend on $a$ and $Z_r$ through the forms:

\begin{equation}
C_1(Z_r,a) = P_1 \left[ X + P_5 Z_r a^3 \right] \ ,
\label{E.18}
\end{equation}
and
\begin{equation}
C_2 = P_2 X,
\label{E.19}
\end{equation}
where

\begin{equation}
X = {(1.0 + P_3 a)}^{P_4} \ ,
\label{E.20}
\end{equation}
\noindent
and the optimum values of the parameters seem to be:

\begin{eqnarray}
P_1 &=&0.1402 \nonumber\\
P_2 &=&0.1285 \nonumber\\
P_3 &=&1.0 \label{E.21} \\
P_4 &=&3.15 \nonumber\\
P_5 &=&4.0 \ . \nonumber
\end{eqnarray}

The chosen forms of the fits are constrained to lie between zero and 1
for all values of $a$ and $Z_r$ and to give the correct functional form
for both small and large limiting values of $\beta$. 
The presented fits have been obtained for data from $a=0$ (no
correlation, which corresponds to the Holtsmark limit) to $a=0.8$.  
For $a$=$Z_r$=0,
the fit is accurate to within $\pm 2 \%$. Except for very small values
of $\beta$, the fit is accurate to within 10$\%$, except for large $a$
and $Z_r$, where it reaches 26$\%$ in the worst case ( for very small
$\beta$). However, as $Q(\beta)$ is very small there ( on the order of
$10^{-4}$ ), it shouldn't matter. 
Figures~\ref{fig1} and~\ref{fig3} 
demonstrate the magnitude of the correlation effects
for a neutral as well as for a singly charged radiator. 

In the following we are going to ignore the temperature and density
dependence of the coefficients $C_1$,$C_2$ in the derivatives of the
$Q(\beta)$. For solar conditions, the maximum relative error of this
approximation is about $10^{-3}$-$10^{-4}$ and, therefore, 
the approximation is quite
sufficient for our fits, which in any case have a few percent
errors themselves. Henceforth,

\begin{equation}
{d f\over{d \beta}} = 
{ 3 C_1 {\beta}^2 ( 1 + {1\over 2} C_2 {\beta}^{3/2}) \over
{ {(1 + C_2 {\beta}^{3/2})}^2} } \ ,
\label{E.22}
\end{equation}

\begin{equation}
{d^2 f\over{d^2 \beta}} = {3\over 4} C_1 \beta
{ ( 8 + 3 C_2 {\beta}^{3/2} + C_2^2 {\beta}^3) \over
{ {(1 + C_2 {\beta}^{3/2})}^3} } \ ,
\label{E.23}
\end{equation}

\begin{equation}
Q^{'}(\beta) = {1\over{ {(1+f)}^2 }} 
{\left({df\over{d\beta}}\right)} \ ,
\label{E.24}
\end{equation}

\begin{equation}
Q^{''}(\beta) = {1\over{ {(1+f)}^2 }} 
                {\left[{
{d^2 f\over{d^2\beta}} - 
{2\over{(1+f)}}{\left({d f\over{d\beta}}\right)}^2
                }\right]} \ .
\label{E.25}
\end{equation}

\section{Q-MHD Equation of State}

\subsection{Q-fit Occupation Probability}

The post-Holtsmark microfield distribution given by the Q-fit can be
used to upgrade the original MHD to the Q-MHD equation of state. Recall
(Section~2.3) that the MHD-style occupation probability is given by a
product of neutral and charged parts

\begin{equation}
w_{ijk} = w_{ijk}^c w_{ijk}^n \ .
\label{E.26}
\end{equation}
\noindent

The neutral part $w_{ijk}^n$, corresponding to pressure
ionization mechanism, has the same form as in the original MHD representation
(see \cite{hm88} 1988). The charged part $w_{ijk}^c$, determined by
the microfield distribution, is given by

\begin{equation}
w_{ijk}^c = Q(\beta_{ijk}) \ ,
\label{E.27}
\end{equation}

\begin{equation}
\beta_{ijk} = { \left( 3\over{4\pi} \right)}^{2/3} {K_{ijk} 
\chi_{ijk}^2 \over{4 (Z_{jk} + 1) e^4}}
{\left( {n_{ion}\over n_e}\right)}^{1/3}{1\over{n_e^{2/3}}}
\ ,
\label{E.28}
\end{equation}
where $\chi_{ijk}$ is the ionization potential of level $i$ of
ion $j$ of chemical species $k$, $Z_{jk}$ is the {\it net} charge on ion
$j$ of species $k$ ( zero for neutral particle), $n_{ion}$ is the total
density of ionic perturbers in a system and $n_e$ is again
the electron density. $K_n$ is the quantal Stark-ionization correction
factor of MHD theory (see \cite{hm88} 1988) 

\begin{eqnarray}
K_{n} &=& 1 \ ,\ \ n\leq 3, \nonumber \\
K_{n} &=& {16\over 3}{n\over {(n+1)}^2} \ ,\ \ n > 3\ .
\label{E.29}
\end{eqnarray}

In the following we adopt an important 
approximation ${\left({n_{ion} / {n_e}}\right)}^{1/3} \approx 1$.  
As been already discussed 
in \cite{hm88} (1988)
the error of this assumption for stellar plasmas of normal chemical
compositions never exceeds a few percent due to scarcity of high-Z
species. This minor approximation simplifies the analytical derivatives
(Appendix) significantly.

Therefore, Equation~(\ref{E.28}) can be rewritten as

\begin{equation}
\beta_{ijk} = \beta_{ijk}(n_e) = 
{ \left( 3\over{4\pi} \right)}^{2/3} 
{K_{ijk} \chi_{ijk}^2 \over{4 (Z_{jk} + 1) e^4}}
{n_e^{-2/3}} \ .
\label{E.30}
\end{equation}

\subsection{Q-MHD Free Energy}

Once the new Q-fit occupation probabilities are known, they merely have
to be put in place of the original MHD occupation probabilities.  
To allow detailed comparisons
with the MHD expressions (\cite{mdh88};~\cite{dmhm88}), we list the
Q-fit analogs in the Appendix. Calculation of the ionization equilibria
and thermodynamic quantities is done as in MHD, and the result is the
Q-MHD equation of state.

\section{Results and Discussion}

\subsection{Comparisons of Q-fit and APEX Microfield Distributions}

\begin{figure}
\figcaption{Microfield distribution from Q-MHD model in a case of
neutral
perturber in hydrogen plasma for different values of
coupling parameter: $\Gamma$ = 0.10 (solid line), $\Gamma$ = 0.25
(dashed line), $\Gamma$ = 1.0 (dotted-dashed line).
\label{fig1}}
\vspace{4cm}
\psfig{file=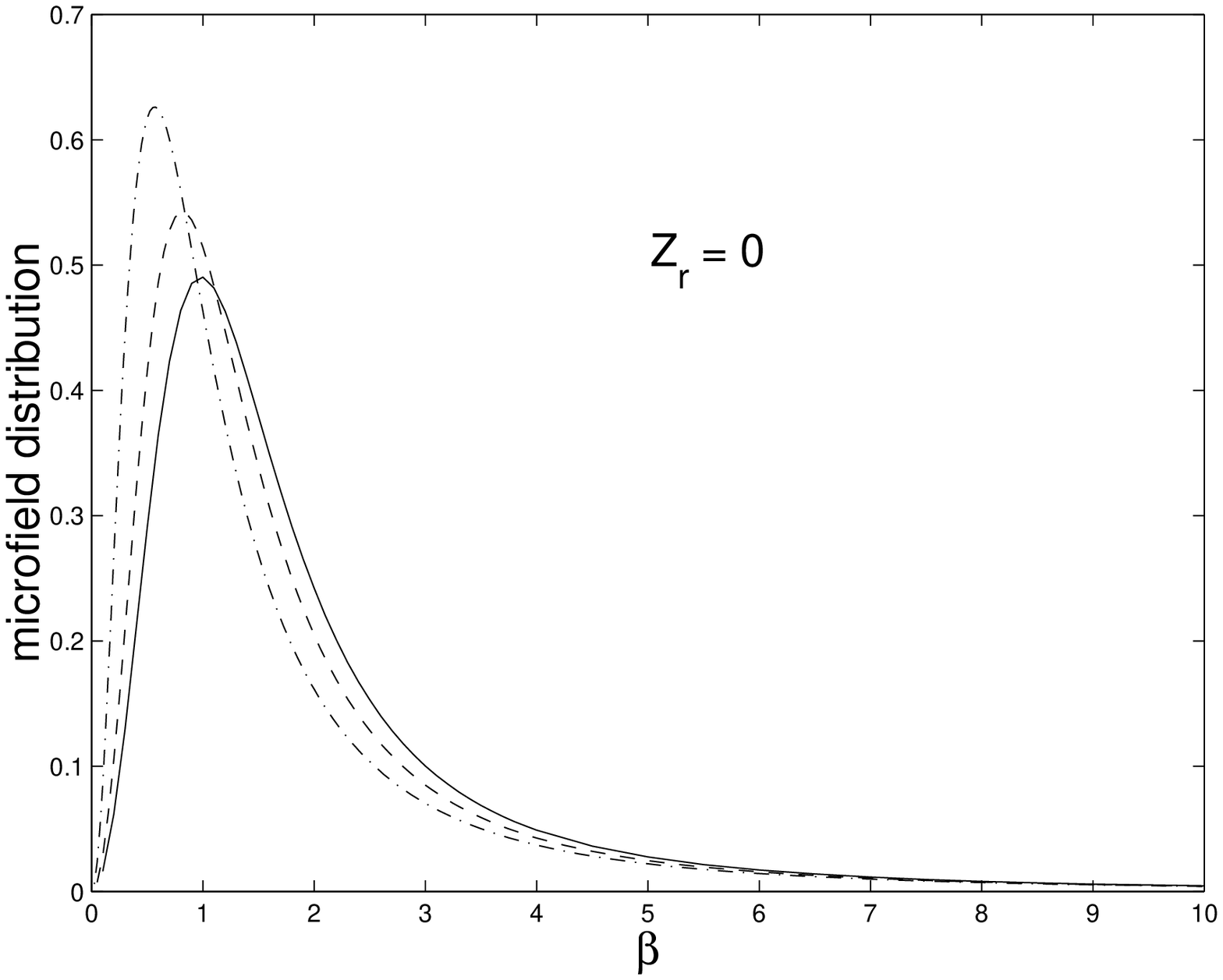}
\end{figure}

\begin{figure}
\figcaption{Microfield distribution from APEX model in a case of neutral
perturber in hydrogen plasma for different values of
coupling parameter: $\Gamma$ = 0.10 (solid line), $\Gamma$ = 0.25
(dashed line), $\Gamma$ = 1.0 (dotted-dashed line).
\label{fig2}}
\vspace{4cm}
\psfig{file=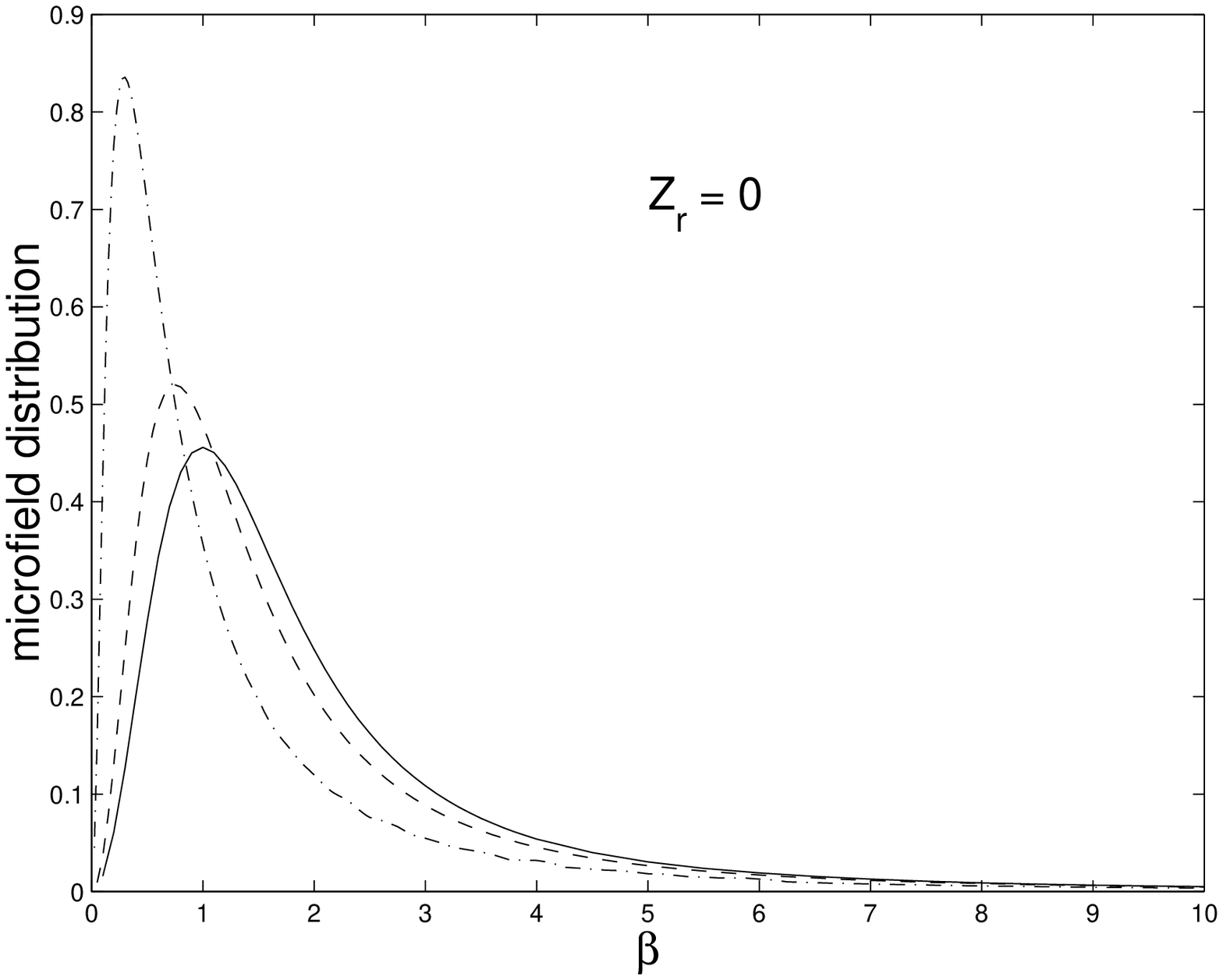}
\end{figure}

\begin{figure}
\figcaption{Same as Fig.1 in a case of $Z_r = 1$ perturber.
\label{fig3}}
\vspace{4cm}
\psfig{file=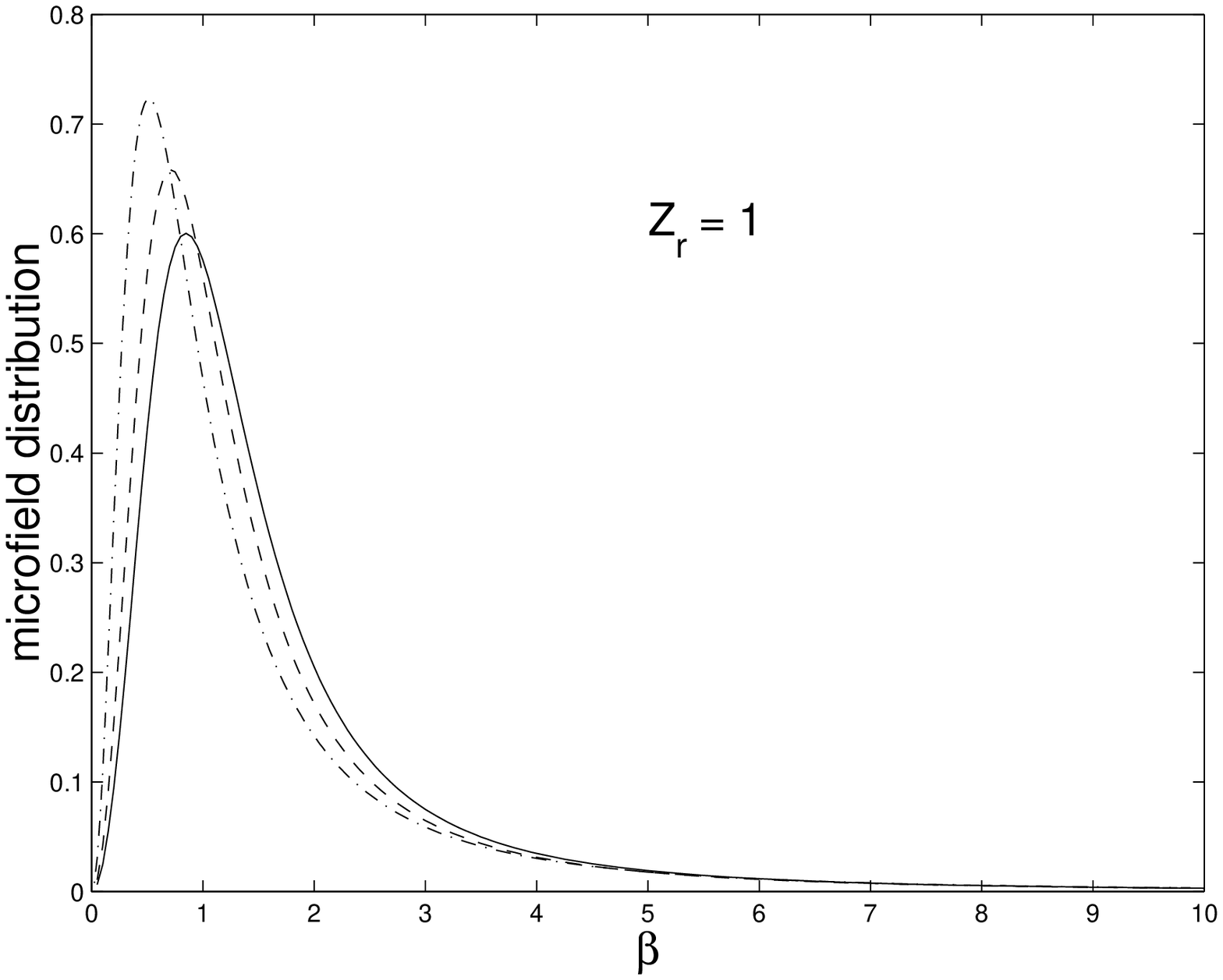}
\end{figure}

\begin{figure}
\figcaption{Same as Fig.2 in a case of $Z_r = 1$ perturber.
\label{fig4}}
\vspace{4cm}
\psfig{file=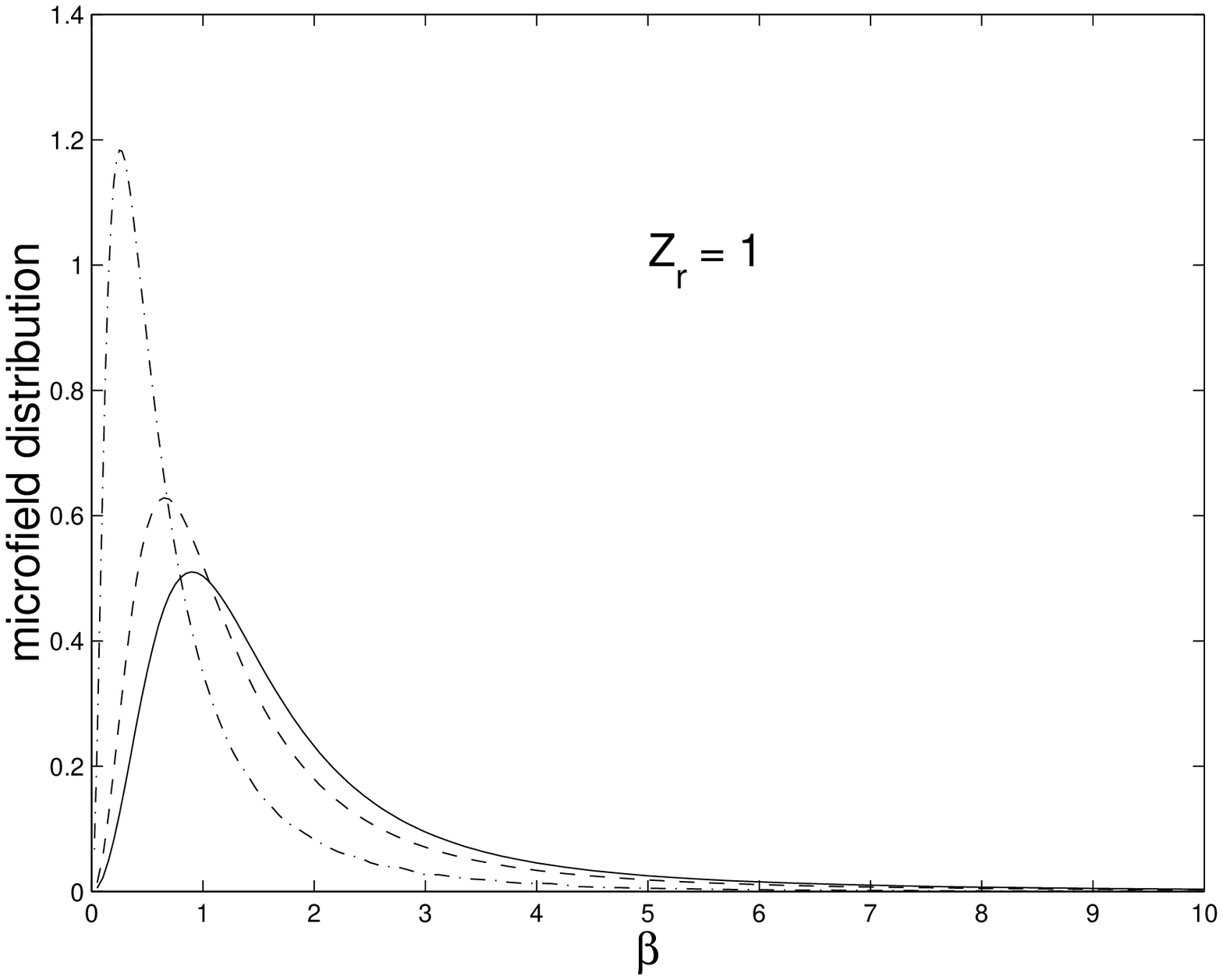}
\end{figure}

In this section the microfield distributions in hydrogen plasmas
are presented for different values of the plasma
coupling parameter $\Gamma$ (which
is the ionic analog to $\Gamma_{\rm e}$ defined by
Eq.~(\ref{E.14}). More precisely,

\begin{equation}
\Gamma = \left({ e^2\over{\kbt a_{\rm av}} }\right) \ ,
\label{E.31}
\end{equation}

\noindent
where $a_{\rm av}$ is the average ion 
sphere radius defined by $({4\pi/3}) \sum_{\rm ion} n_i a_{\rm av}^3 =1$.
In both Q-MHD and APEX formalisms only singly-charged perturbers
considered. To demonstrate the effect of a radiator charge, the case
of a neutral ($Z_r = 0$) radiator (see Figs.~\ref{fig1}
and~\ref{fig2}) is 
presented alongside with the singly-charged ($Z_r = 1$) case 
(Figs.~\ref{fig3}-\ref{fig4})
calculated for a several values of coupling parameter $\Gamma$.

\placefigure{fig1}
\placefigure{fig2}
\placefigure{fig3}
\placefigure{fig4}

The resultant function Q [see Eq.~(\ref{E.9})] for these cases
demonstrates a clear dependence on a radiator charge as well as on a
value of $\Gamma$. 

\begin{figure}
\figcaption{Function Q for a case presented in Fig.1
\label{fig5}}
\vspace{4cm}
\psfig{file=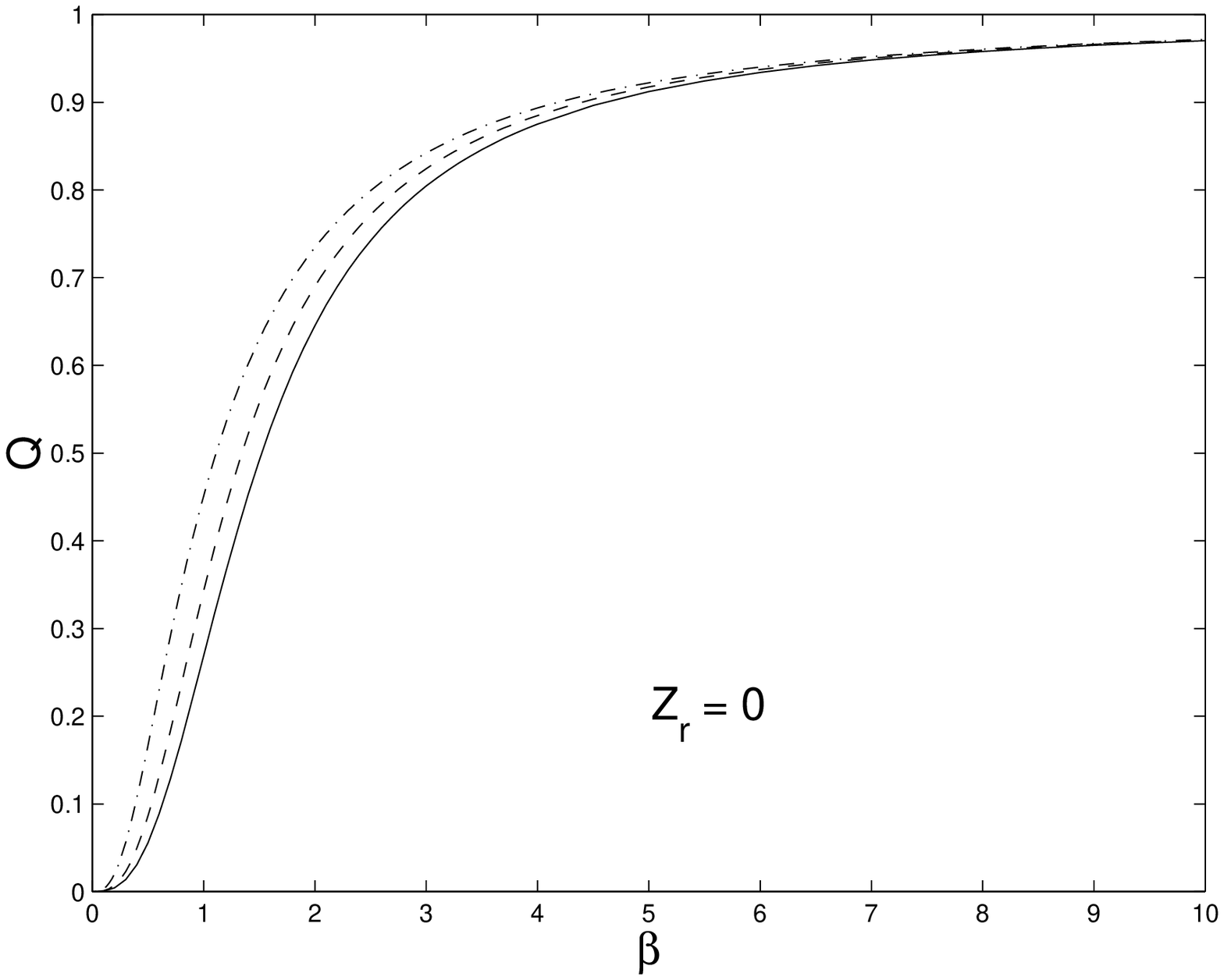}
\end{figure}

\begin{figure}
\figcaption{Function Q for a case presented in Fig.2
\label{fig6}}
\vspace{4cm}
\psfig{file=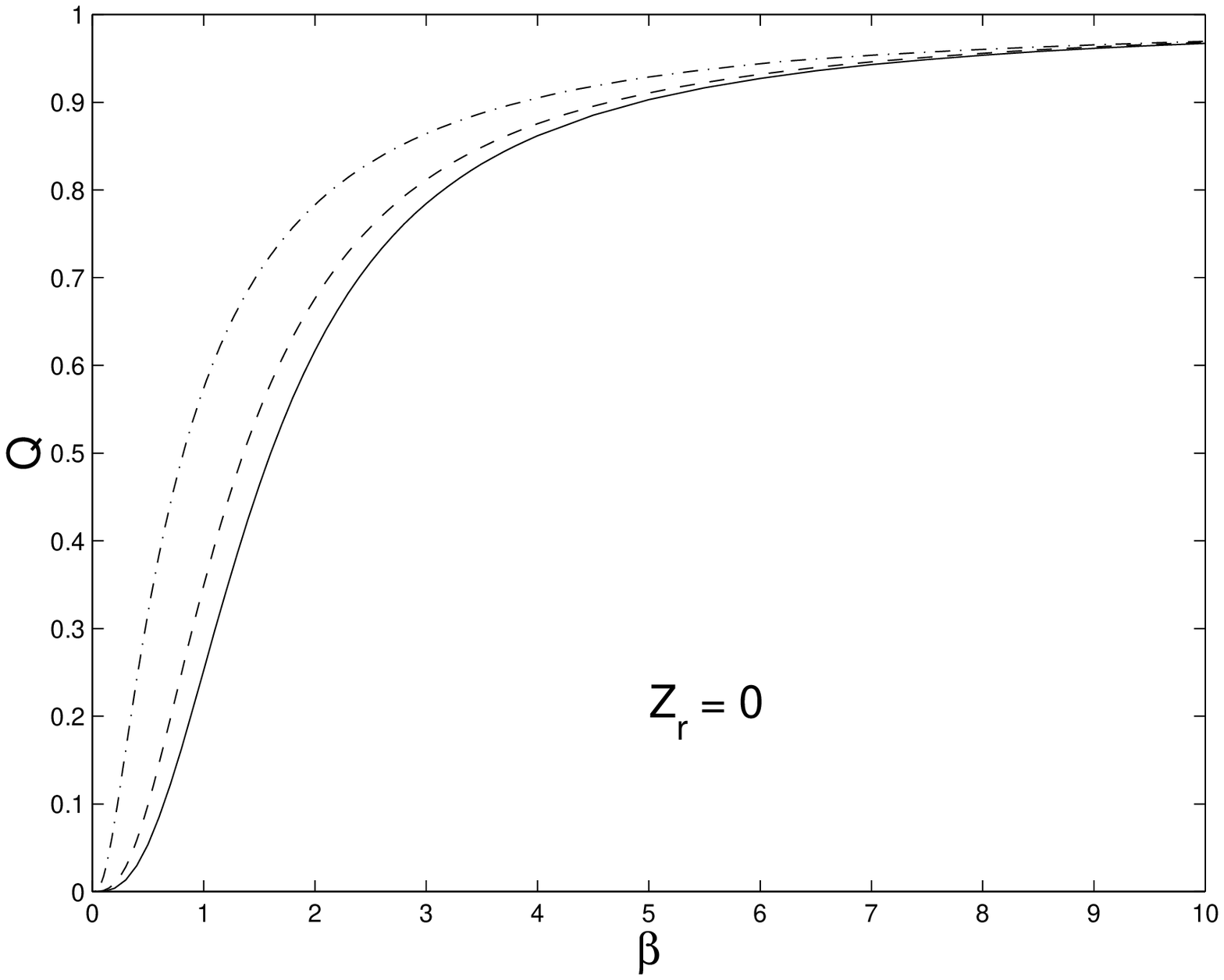}
\end{figure}

\begin{figure}
\figcaption{Function Q for a case presented in Fig.3
\label{fig7}}
\vspace{4cm}
\psfig{file=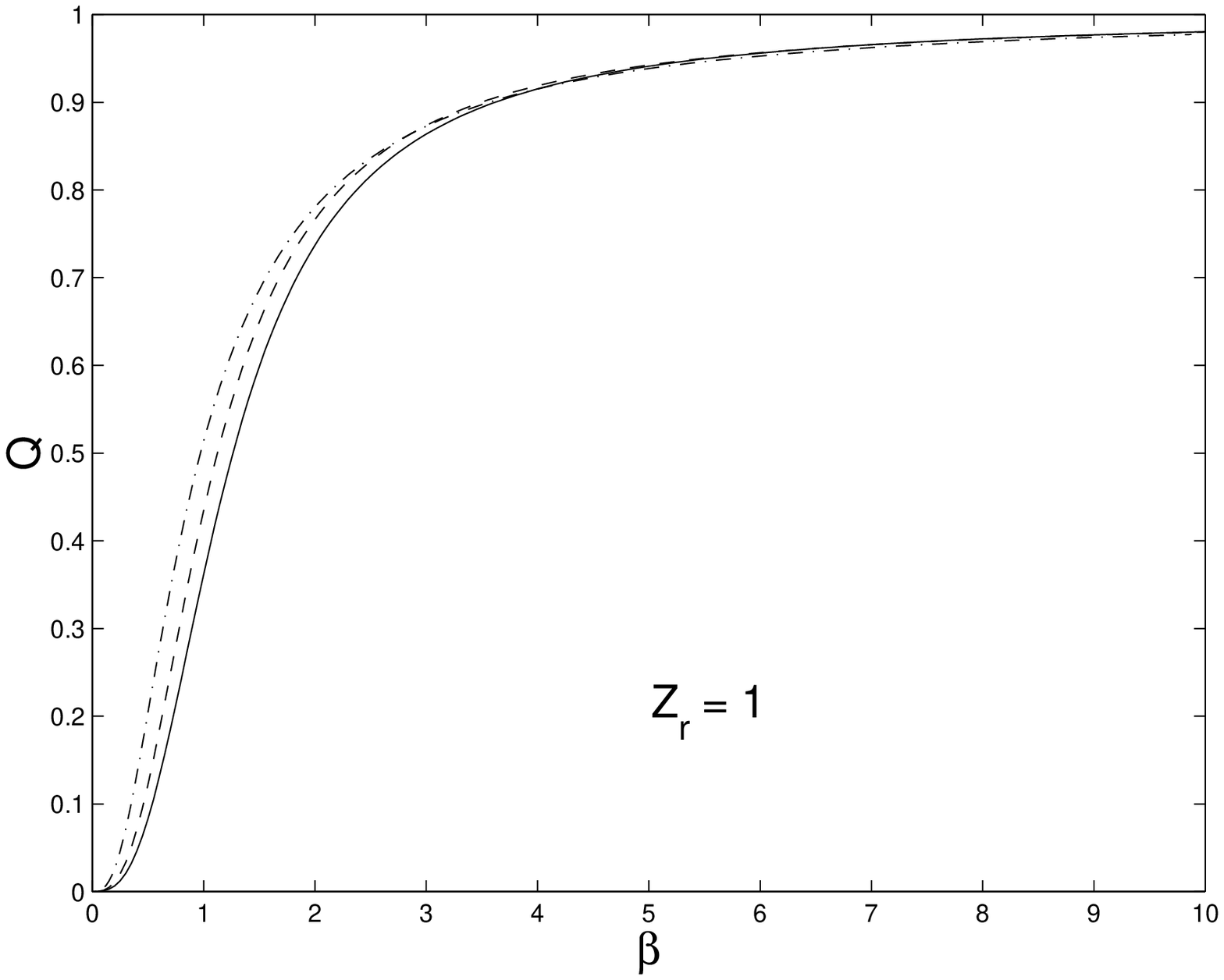}
\end{figure}

\begin{figure}
\figcaption{Function Q for a case presented in Fig.4
\label{fig8}}
\vspace{4cm}
\psfig{file=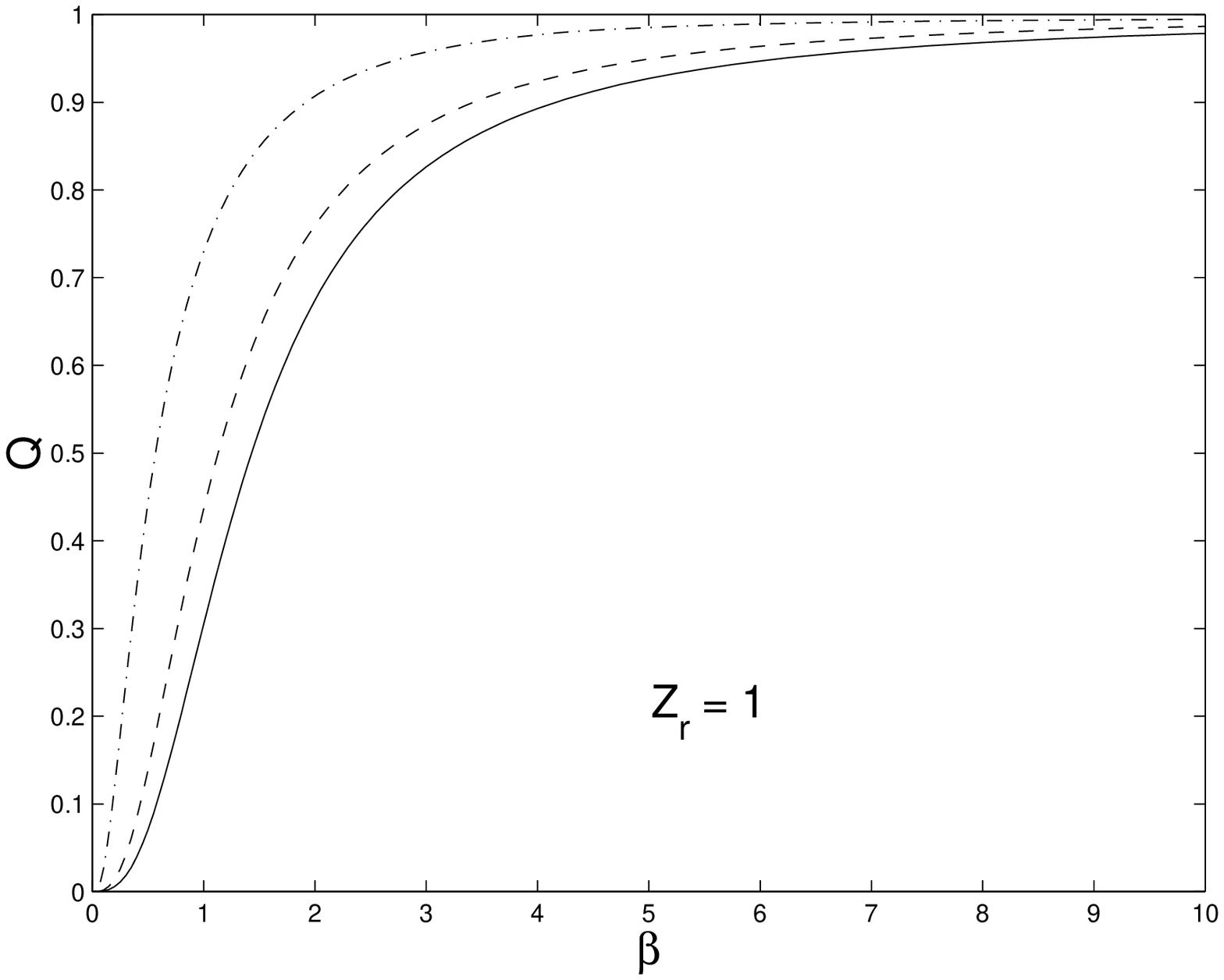}
\end{figure}

\placefigure{fig5}
\placefigure{fig6}
\placefigure{fig7}
\placefigure{fig8}

\subsection{Comparison of the Equation of State for a Hydrogen-only Plasma}

\begin{figure}
\figcaption{
``H-only solar track'': for given
temperature, density of a hydrogen-only plasma is chosen such that 
pressure corresponds to solar pressure.
\label{fig9}}
\vspace{4cm}
\psfig{file=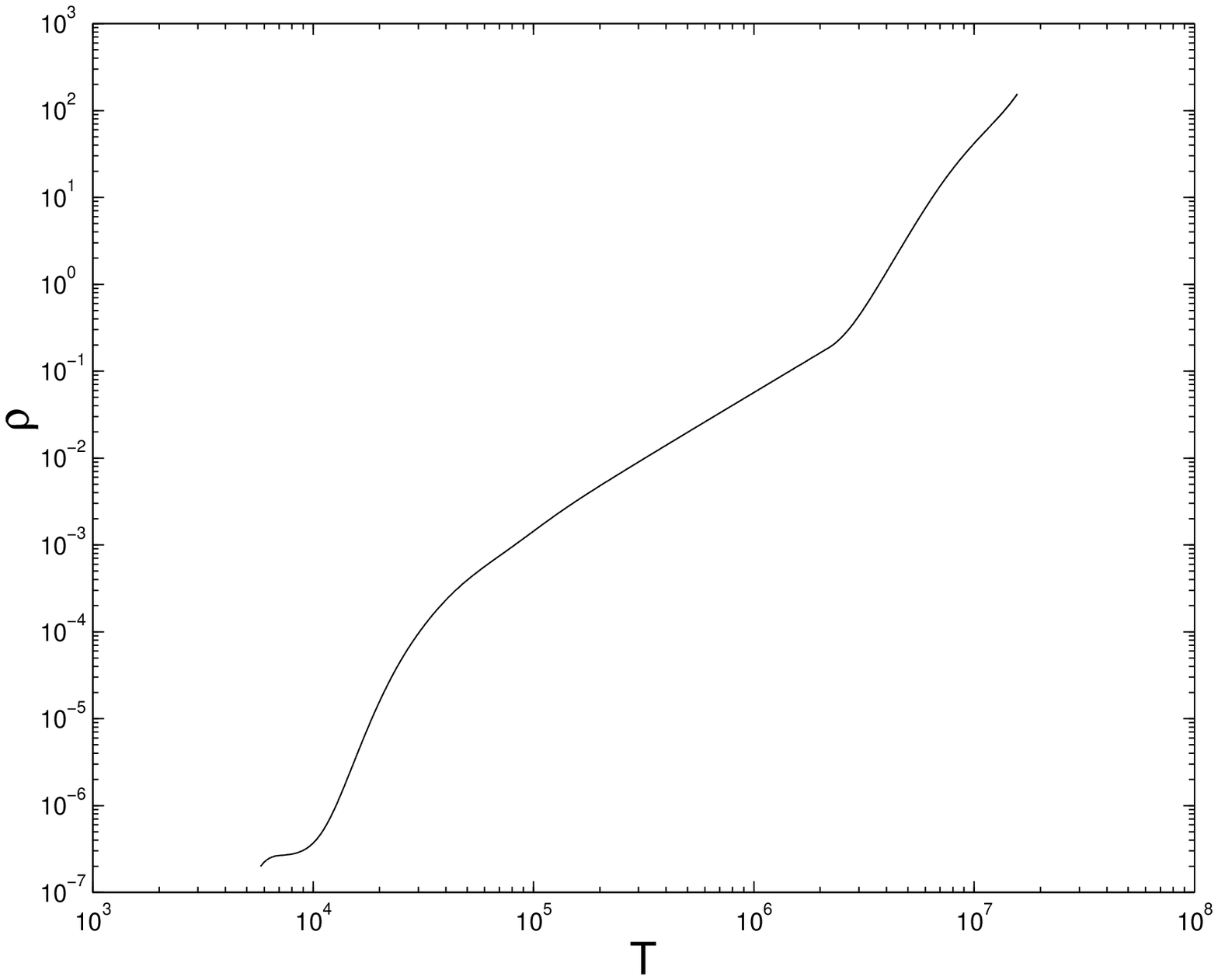}
\end{figure}

\begin{figure}
\psfig{file=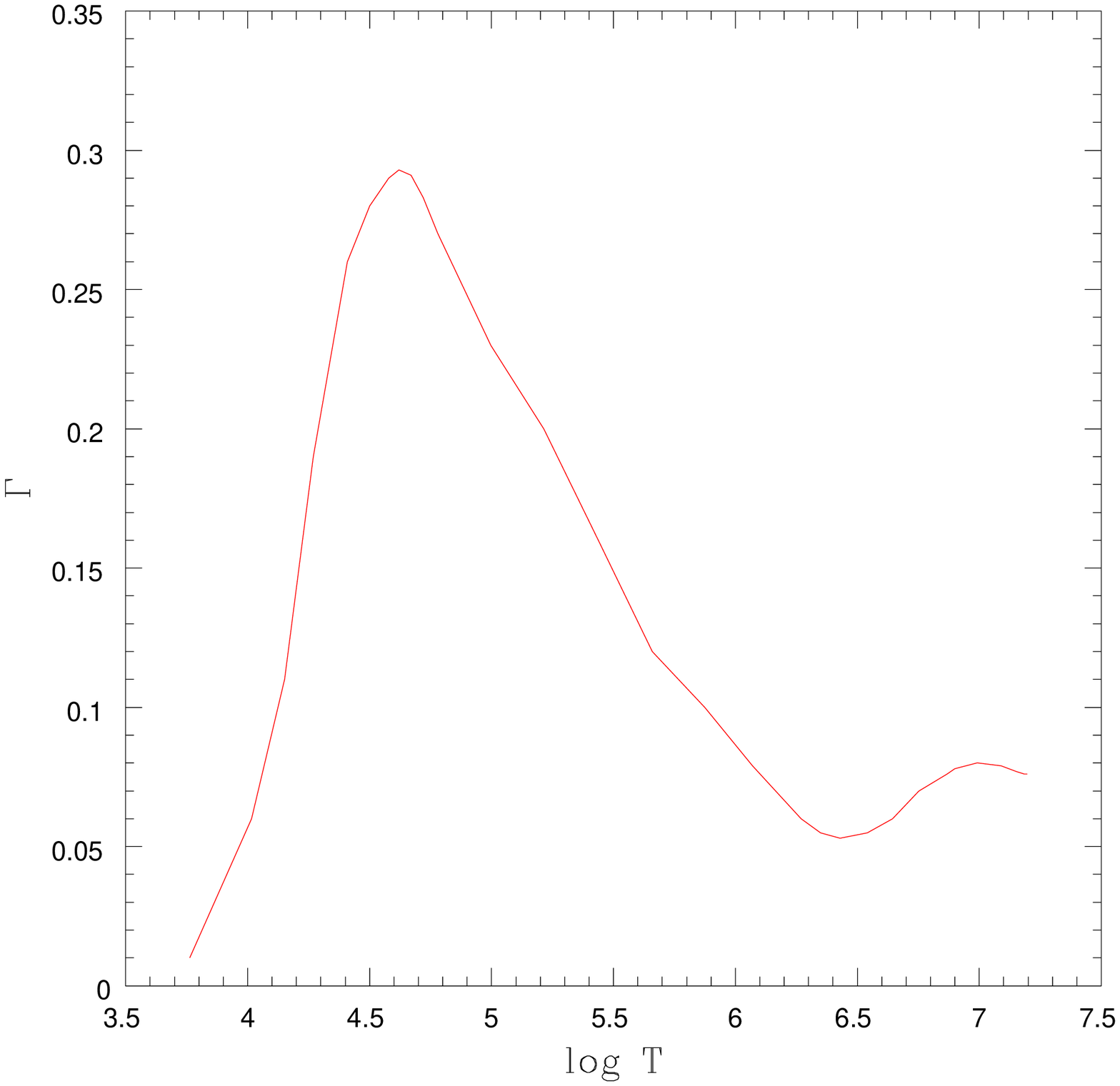}
\vspace{4cm}
\figcaption{Estimated coupling parameter $\Gamma$ along the H-only
solar track 
\label{fig10}}
\end{figure}

This section deals with equation-of-state
calculations for hydrogen-only plasmas.
Four different equations of state
have been calculated for a set
of fictitious solar temperatures and densities, given in
Fig.~\ref{fig9}. 
The densities were
chosen by~\cite{nd98} (1998) 
to simulate solar pressure at the given temperature for a
hydrogen-only plasma. In the following we will denote by ``H-only solar
track'' the set of temperatures and densities of
Figure~\ref{fig9}. Figure~\ref{fig10}
demonstrates a coupling parameter $\Gamma$ as estimated for the
conditions of the H-only solar track.

\placefigure{fig9}
\placefigure{fig10}

\begin{figure}
\psfig{file=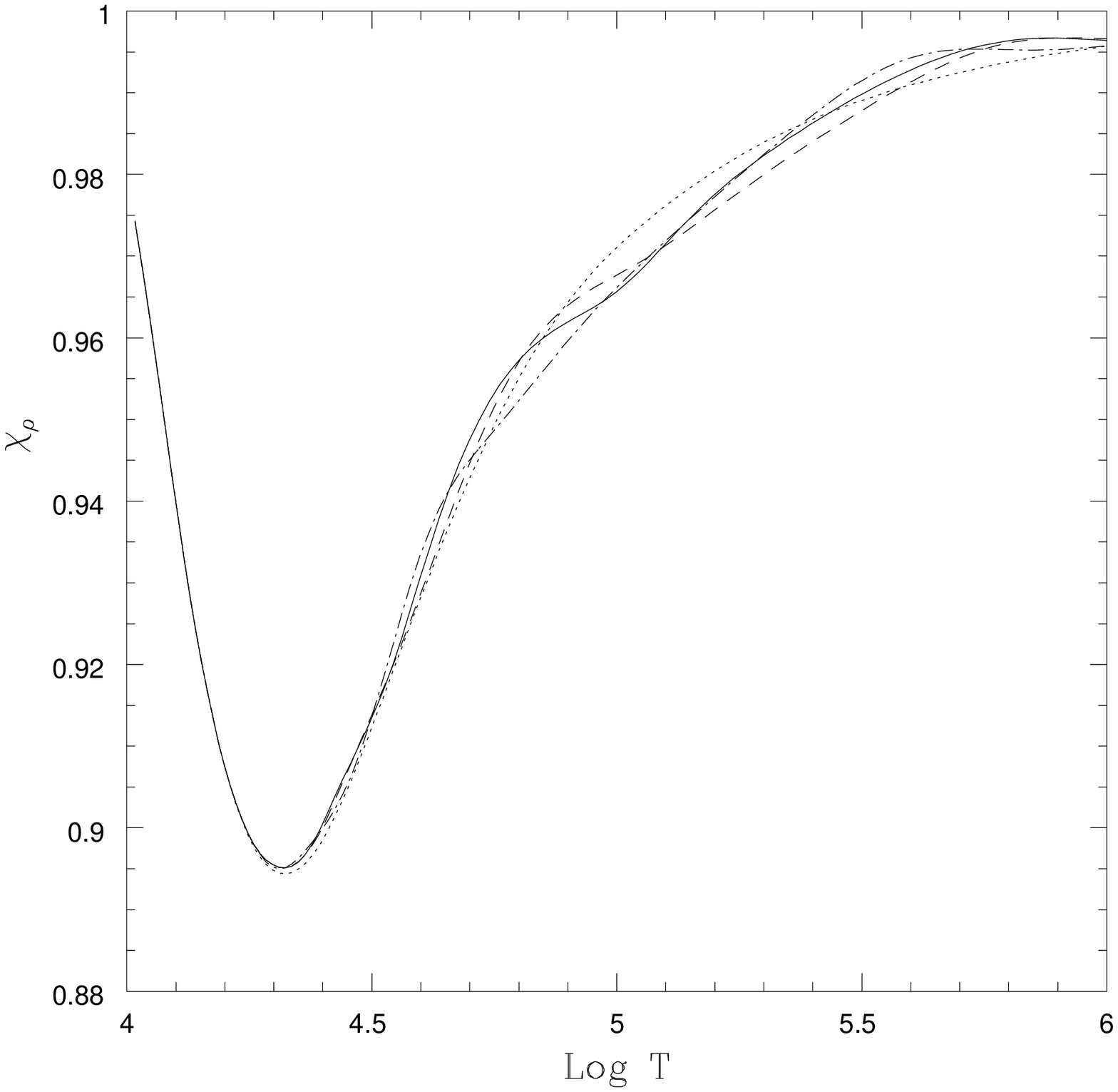}
\vspace{2cm}
\figcaption{$\chi_{\rho}$ for a case of hydrogen-only plasma calculated
for 4 different models: standard MHD (solid line), MHD with Holtzmark
(dotted-dashed line), Q-MHD (dashed line) and OPAL (dotted line). For
details see text.
\label{fig11}}
\end{figure}

\begin{figure}
\psfig{file=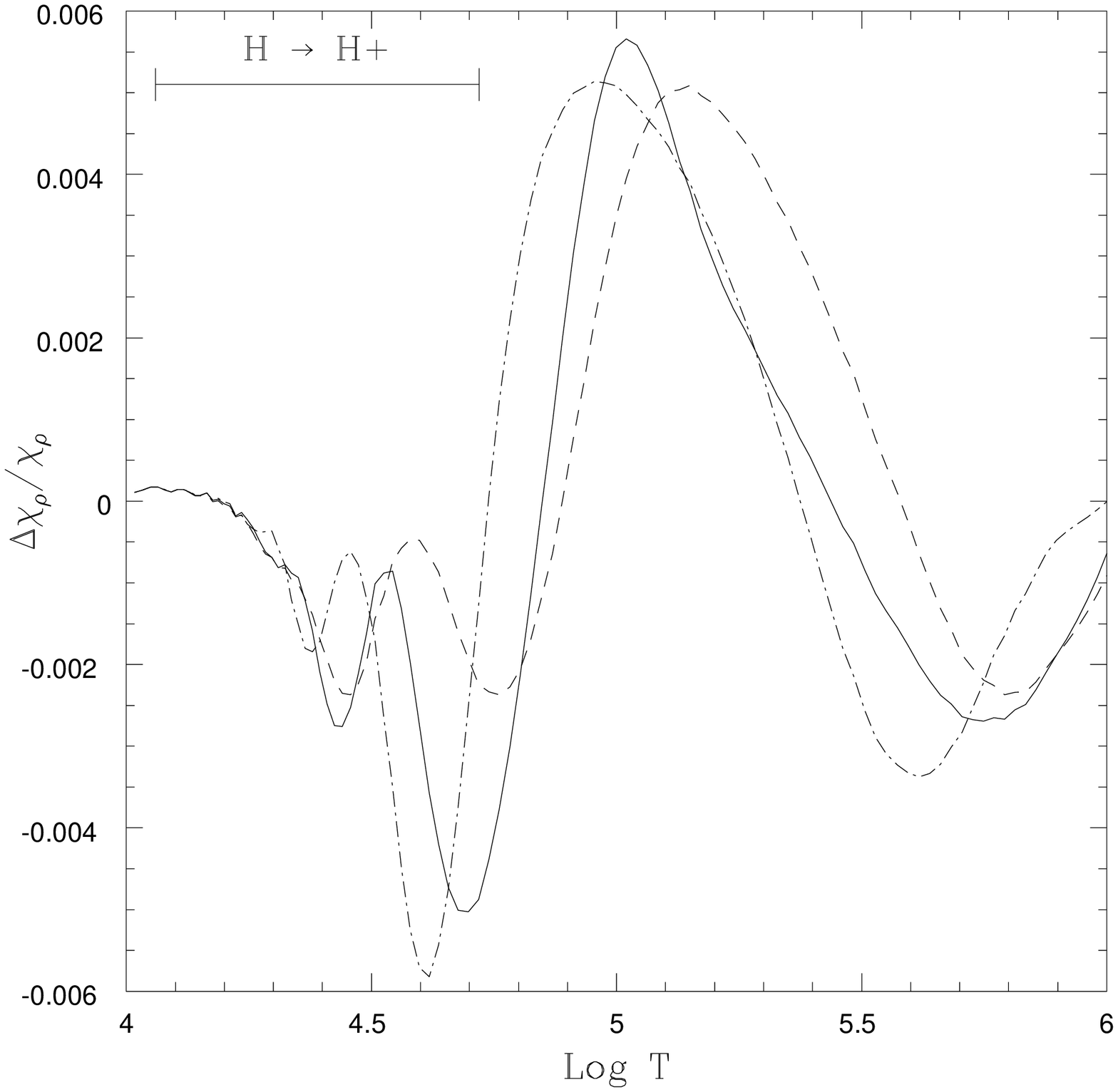}
\vspace{2cm}
\figcaption{
Relative differences in $\chi_{\rho}$ with respect to OPAL 
for a case of hydrogen-only plasma. Linestyles are the same
as in Fig.~\ref{fig11}. The ionization zone of hydrogen 
(ionization degree between 10\% and 90\%) is
indicated. For details see text.
\label{fig12}}
\end{figure}

The four equations of state (EOS) are:  
(i) regular MHD EOS (solid lines 
in all of the following graphs); 
(ii) MHD EOS with a true Holtsmark
microfield distribution function (dotted-dashed lines); (iii)
Q-MHD EOS, as described in Section~5 of this paper (dashed lines) and,
finally, (iv) OPAL EOS (dotted lines).  The results
expand the previous study (\cite{nd98} 1998) 
on the effects of internal
partition functions, and in as much as the studies
overlap, they agree with each other. 
Among the quantities considered, that is,
$\chi_{\rho} = {(\partial \ln p / \partial \ln \rho)}_T$, 
$\chi_T =  {(\partial \ln p / \partial \ln T)}_{\rho}$, 
and $\gamma_1 = {(\partial \ln p / \partial \ln \rho)}_s$, 
only $\chi_{\rho}$ reveals differences already in the
{\it absolute} plot. However, the differences are
present, and of the same order of magnitude, also in the
other two quantities, in the same way as found by
\cite{nd98} (1998).
Figures~\ref{fig11} and~\ref{fig12} show absolute
values of $\chi_{\rho}$
and relative difference (with respect to OPAL), respectively. 
All three MHD-type EOS ({\it i.e.},
MHD, MHD with true Holtsmark, and Q-MHD) demonstrate 
the characteristic
wiggle discovered in the previous study (see also
Fig.~\ref{fig-hel2}). The wiggle
is located in the temperature zone $4.5 \le \log T
\le 5.3$. Given the hydrogen-only plasma, 
it is obviously a pure hydrogen phenomenon. 
However, as the location of the 10\% to 90\% ionization zone
demonstrates (Fig.~\ref{fig12}), the wiggle is
caused by the last remaining neutral hydrogen atoms,
fighting against pressure ionization in a region of near
full ionization.

\placefigure{fig11}
\placefigure{fig12}

The figures also reveal a close agreement between
regular MHD and MHD with a true Holtsmark distribution function,
which gives a {\it post factum} justification of the
choice made by the authors of MHD in 1988.  It is also clear that Q-MHD
seems to be in a better overall agreement with OPAL than other two
equations of state.

\subsection{Comparison of the Equation of State for a Hydrogen-Helium Mixture}

\begin{figure}
\psfig{file=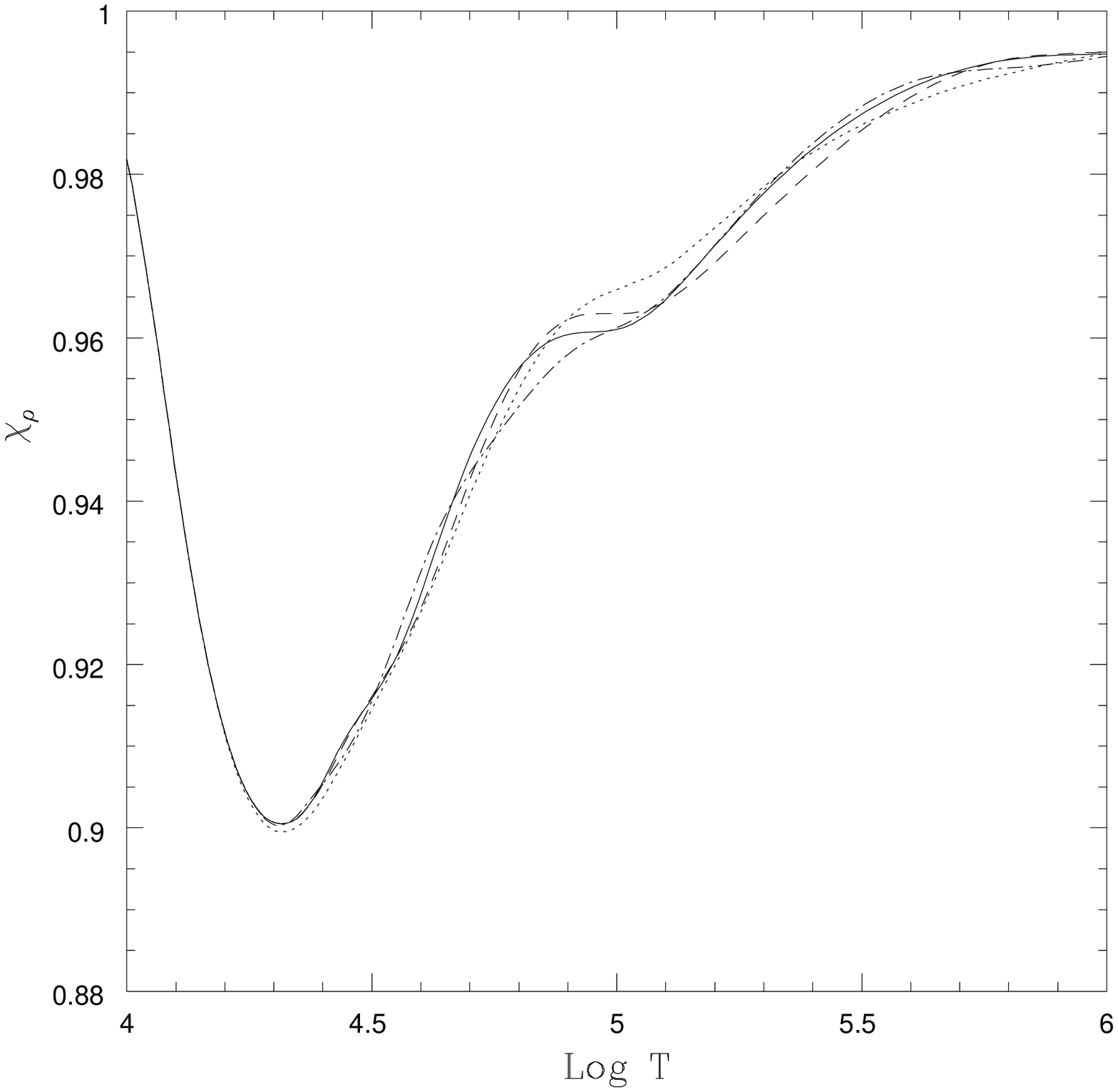}
\vspace{2cm}
\figcaption{
The same as Fig.~\ref{fig11} 
for a case of a hydrogen-helium mixture (74\% to 26\% by
mass, respectively).
For details see text.
\label{fig13}}
\end{figure}

\begin{figure}
\psfig{file=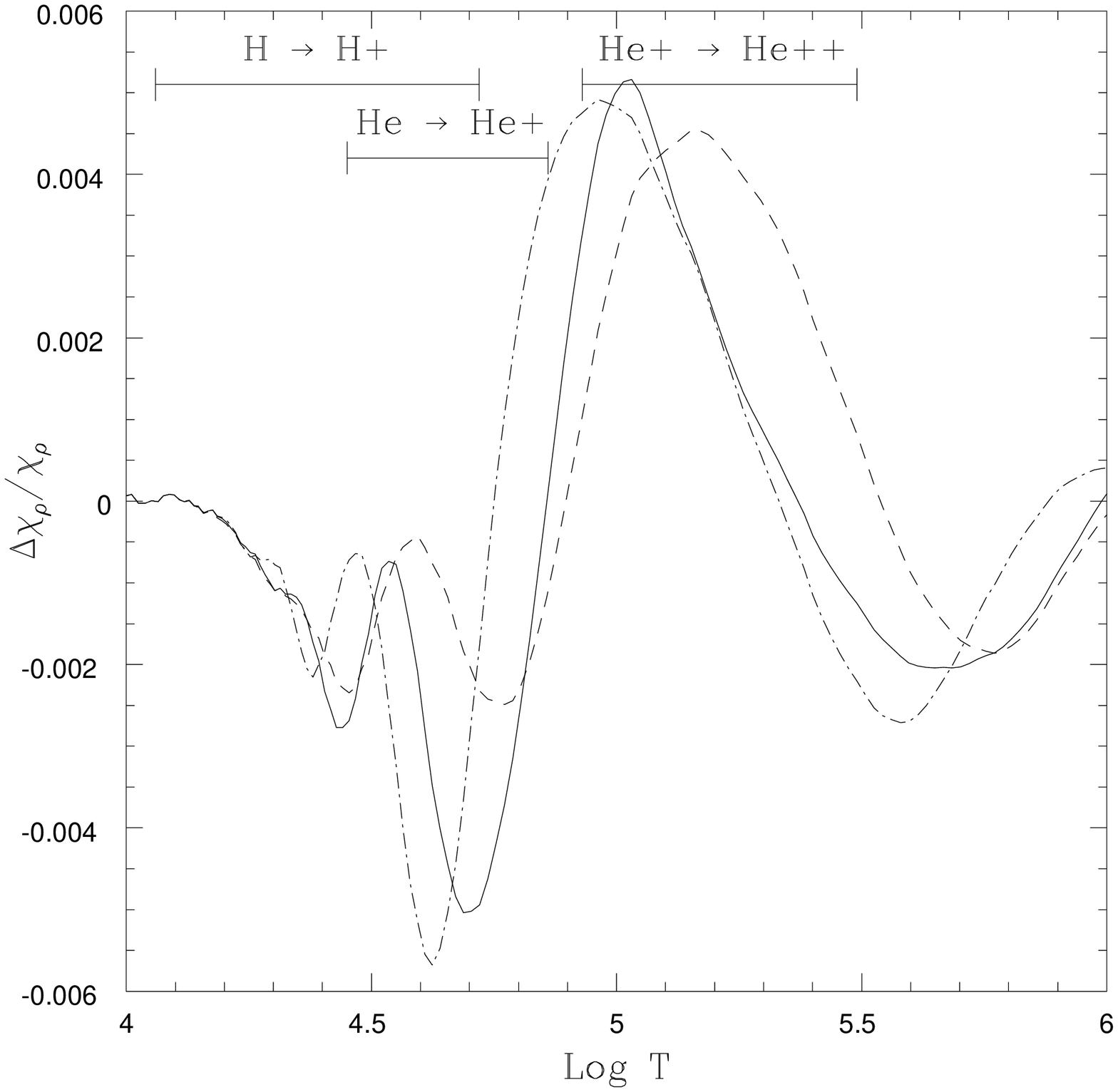}
\vspace{2cm}
\figcaption{
The same as Fig.~\ref{fig12} 
for the hydrogen-helium mixture of Fig.~\ref{fig11}. 
Besides the hydrogen ionization zone, the two ionization zones of 
helium (ionization degree between 10\% and 90\%) are also indicated.
For details see text.
\label{fig14}}
\end{figure}

To study a more realistic picture of solar plasmas, calculations similar
to ones described in the previous section were carried out for a
hydrogen-helium mixture along a real solar profile of density and
temperature (model S of~\cite{cd96}). Hydrogen constitutes 74\% of this
mixture by mass, making it very similar to a solar composition of the
standard solar model. While those calculations exclude heavier elements
from the analysis, the obtained results actually give a strong
indication that a treatment of heavier elements is necessary
to improve our current equation of state models (\cite{da93}).  The
plots of absolute values of $\chi_{\rho}$ and of relative differences
(with respect to OPAL)
between the same four equations of state are given by
Figures~\ref{fig13} and~\ref{fig14}.

\begin{figure}
\psfig{file=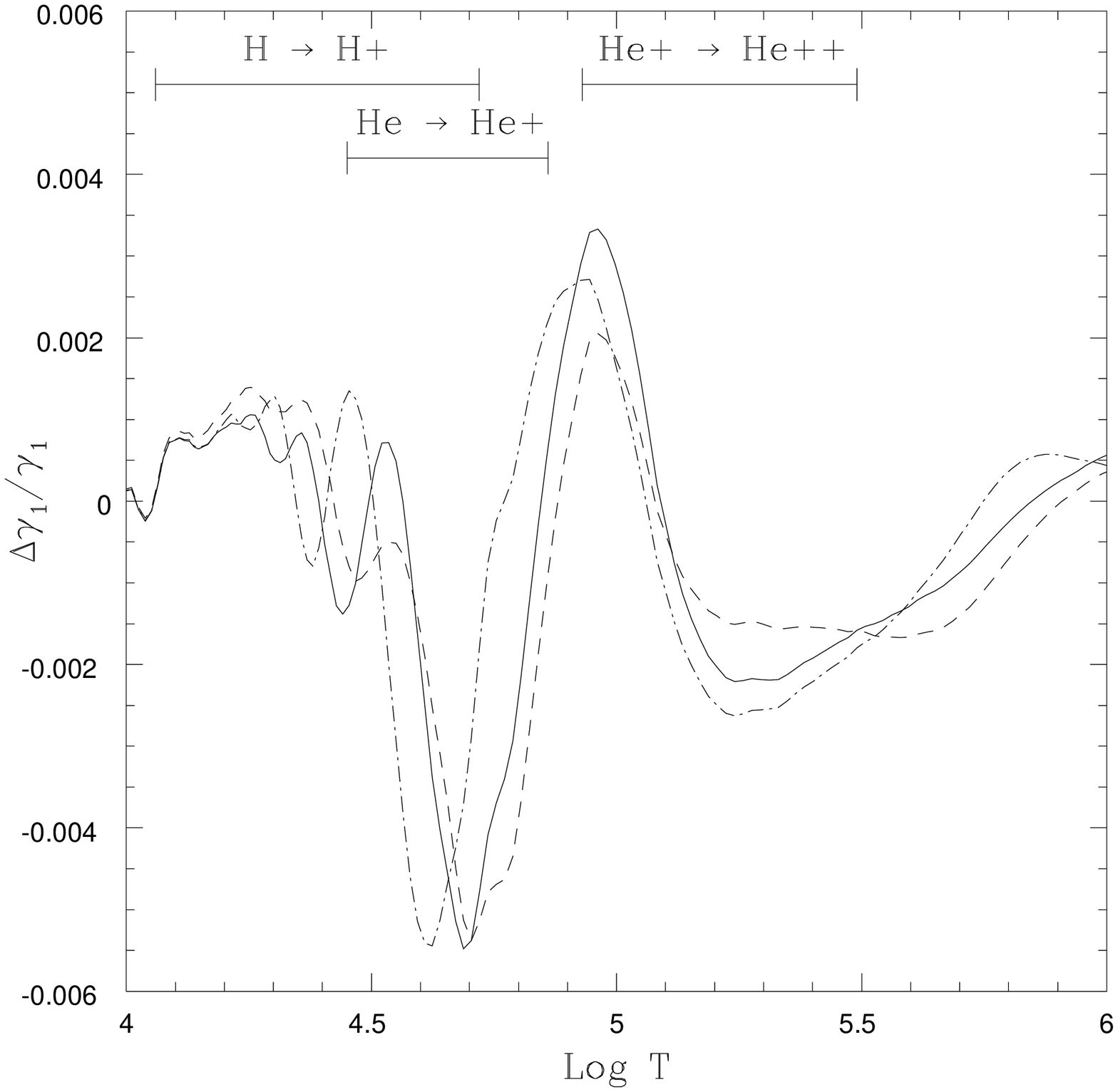}
\vspace{2cm}
\figcaption{
Relative differences in the adiabatic index
$\gamma_1$ with respect to OPAL 
for the hydrogen-helium mixture of Fig.~\ref{fig11}. 
Ionization degrees are as
in Fig.~\ref{fig14}, linestyles as in Fig.~\ref{fig11}.
For details see text.
\label{fig15}}
\end{figure}

\placefigure{fig13}
\placefigure{fig14}
\placefigure{fig15}

Figure~\ref{fig15} 
reveals that the signature of the hydrogenic internal partition
function is present in other thermodynamic quantities,
such as $\gamma_1$, as well. 
Again, one can argue that Q-MHD agrees with OPAL better than the
other
models and that the differences between the usual MHD and MHD with a true
Holtsmark
distribution remain small.
Fig.~\ref{fig15} shows that
Q-MHD leaves MHD and approaches OPAL especially in the
temperature range $5.1 \le \log T \le 5.4$. From the
comparison with the helioseismic experiment
(Fig.~\ref{fig-hel3}), one realizes that for at least the
upper part of
this temperature range, 
OPAL is a better equation of state than MHD. Therefore, Q-MHD
corrects MHD in the right direction.

\section{Conclusion}

Upgrading the MHD equation of state to include realistic microfield
distributions beyond the Holtsmark approximation has confirmed the
significant changes in the occupation numbers for atomic and ionic
states, as was expected by \cite{ir95} (1995). 
These changes in the occupation numbers
and the associated shifts in the ionization balances are widely assumed
responsible, among other, for the discrepancies between OP and OPAL
opacities 
under the temperatures and densities of
the solar center (see~\cite{gdl98}). For conditions of
envelopes
of more massive stars, however, the two opacity calculations
agree very well. Our study is a systematic
comparison of the impact on occupation probabilities of the
original MHD microfield, the proper Holtsmark
microfield, and the APEX distribution. 

As far as {\it thermodynamic} properties are concerned, 
\cite{ir95} (1995) and also
\cite{ri98} (1998) 
believed that effects of different microfield distributions
would be rather negligible. But encouraged by recent progress on the
influence of excited states in hydrogen on thermodynamic quantities
(Section~3),
we have found a clear signature of the microfield distribution,
easily within reach of helioseismological accuracy
(compare Figures~\ref{fig-hel3} and~\ref{fig15}). This
confirms once more that solar
observations constrain formalisms used to
describe the physics of atoms and compound ions immersed in a plasma.

Since these thermodynamic effects are nonetheless quite small and are,
even for helioseismological accuracy, only relevant at some selected
locations of the Sun, our results also show that for most applications
of stellar structure, the approximations chosen in the original MHD
equation of state are reasonable. In the particular cases where
they are not reasonable, such as in
helioseismological studies of 
the zones of partial ionization in the Sun, the more
accurate Q-MHD equation of state, which is 
based on a realistic microfield
distribution, is more accurate.

\section{Acknowledgments}

We thank J\o rgen Christensen--Dalsgaard, Carlos Iglesias and
Forrest Rogers for stimulating discussions and 
critical comments. We are grateful to Sarbani Basu for providing 
Figures~\ref{fig-hel1} and~\ref{fig-hel3}. A.N.
and W.D. are
supported by the grant AST-9618549 of the National Science
Foundation. W.D. acknowledges additional support from
the SOHO Guest Investigator
Grant NAG5-7352 of NASA,
the Danish National Research
Foundation through its establishment of the Theoretical Astrophysics Center,
and from PPARC.
D.M. was supported by the Mc Vittie Professorship Fund at the
University of Illinois during the early phases of this work.
SOHO
is a project of international cooperation between ESA and NASA.

\appendix

\section{Appendix: Analytical Expressions for the Free Energy and Its
Derivatives with Q-fit Occupation Probabilities}

We follow the MHD notation 
(\cite{mdh88};~\cite{dmhm88}) very closely. 
The term affected by the new microfield is the free energy of
bound systems of species $s$

\begin{equation}
F_2 = \sum_{s \neq e}N_{s}\left(E_{1s} -\kbt \log Z_s \right) \ ,
\label{W.2.14}
\end{equation}
with the internal partition function evaluated for energies relative
to the ground state, that is,

\begin{equation}
Z_s = 
\sum_i w_{is} g_{is} \exp \left( {-\Delta E_{is}\over{\kbt}} \right) \ .
\label{W.2.13}
\end{equation}
Here, $\Delta E_{is} = E_{is} - E_{1s}$.
In what follows, $\lambda, \mu, \nu$ label neutral, 
$q$ charged particles. We shall also
treat ($N_p/N_e)$ as a constant (see the discussion leading to
Eq.~(\ref{E.30}). Then, (1) $F_2$ is linear in $N_{j}$, 
and (2), $Z_s$ does not depend on individual ions $N_j$. Then

\begin{equation}
{\partial F_2\over{\partial N_\nu}} = 
E_{1\nu} - \kbt \ln Z_{\nu} - \kbt \sum_{s\neq e}{N_s\over Z_s} 
{\partial Z_s\over{\partial N_{\nu}}} 
\label{A.0.1}
\end{equation}

\begin{equation}
{\partial F_2\over{\partial N_q}} = E_{1q} - \kbt \ln Z_q
\label{A.0.2}
\end{equation}

\begin{equation}
{\partial F_2\over{\partial N_e}} = - \kbt \sum_{s\neq e}{N_s\over Z_s} 
{\partial Z_s\over{\partial N_e}} 
\label{A.0.3}
\end{equation}

\begin{equation}
{\partial^2 F_2\over{\partial N_e^2}} = - \kbt \sum_{s\neq e} {N_s\over Z_s}
{\left[{ {\partial^2 Z_s\over{\partial N_e^2}} - {1\over{Z_s}}
{\left({\partial Z_s\over{\partial N_e}}\right)}^2}\right]}
\label{A.0.4}
\end{equation}

\begin{equation}
{\partial^2 F_2\over{\partial N_e \partial N_q}} = - \kbt 
{1\over{Z_q}} {\partial Z_q\over{\partial N_e}}
\label{A.0.5}
\end{equation}

\begin{equation}
{\partial^2 F_2 \over{\partial N_\lambda \partial N_e} }
= - \kbt 
\left[
{1\over{Z_\lambda}} {\partial Z_\lambda\over{\partial N_e}} + 
\sum_{s\neq e} {N_s\over Z_s}
\left( 
{\partial^2 Z_s\over{\partial N_e \partial N_\lambda}} - {1\over{Z_s}}
{{\partial Z_s\over{\partial N_e}}}
{{\partial Z_s\over{\partial N_\lambda}}}
\right) \right]
\label{A.0.6}
\end{equation}

\begin{equation}
{\partial^2 F_2 \over{\partial N_\lambda \partial N_\mu} }
= - \kbt 
\left[
{1\over{Z_\lambda}} {\partial Z_\lambda\over{\partial N_\mu}} + 
{1\over{Z_\mu}} {\partial Z_\mu\over{\partial N_\lambda}} + 
\sum_{s\neq e} {N_s\over Z_s}
\left( 
{\partial^2 Z_s\over{\partial N_\mu \partial N_\lambda}} - {1\over{Z_s}}
{\partial Z_s\over{\partial N_\mu}}
{\partial Z_s\over{\partial N_\lambda}}
\right)
\right]
\label{A.0.7}
\end{equation}

\begin{equation}
{\partial^2 F_2 \over{\partial T \partial N_e} }
= - \kbt 
\sum_{s\neq e} {N_s\over Z_s}
{\left[{ 
{\partial^2 Z_s\over{\partial N_e \partial T}} +
{\left({ {1\over T} - {1\over Z_s}{\partial Z_s\over{\partial T}}}\right)}
{\partial Z_s\over{\partial N_e}}
}\right]}
\label{A.0.8}
\end{equation}

\begin{equation}
{\partial^2 F_2 \over{\partial T \partial N_\lambda} }
= - \kbt 
{\left\lbrace{
{1\over T} \ln Z_\lambda +
{1\over{Z_\lambda}} {\partial Z_\lambda\over{\partial T}} + 
\sum_{s\neq e} {N_s\over Z_s}
{\left[{ 
{\partial^2 Z_s\over{\partial T \partial N_\lambda}} +
{\left({ {1\over T} - {1\over Z_s}{\partial Z_s\over{\partial T}}}\right)}
{\partial Z_s\over{\partial N_\lambda}}
}\right]}
}\right\rbrace}
\label{A.0.9}
\end{equation}

\begin{equation}
{\partial^2 F_2 \over{\partial T \partial N_q} }
= - \kbt 
\left(
{1\over T} \ln Z_q +
{1\over{Z_q}} {\partial Z_q\over{\partial T}} 
\right)
\label{A.0.10}
\end{equation}

\begin{equation}
{\partial^2 F_2 \over{\partial V \partial N_\lambda} }
= - \kbt 
\left[
{1\over{Z_\lambda}} {\partial Z_\lambda\over{\partial V}} + 
\sum_{s\neq e} {N_s\over Z_s}
\left( 
{\partial^2 Z_s\over{\partial V \partial N_\lambda}} - {1\over{Z_s}}
{\partial Z_s\over{\partial V}}
{\partial Z_s\over{\partial N_\lambda}}
\right)
\right]
\label{A.0.11}
\end{equation}

\begin{equation}
{\partial^2 F_2 \over{\partial V \partial N_e} }
= - \kbt 
\sum_{s\neq e} {N_s\over Z_s}
\left( 
{\partial^2 Z_s\over{\partial V \partial N_e}} - {1\over{Z_s}}
{\partial Z_s\over{\partial V}}
{\partial Z_s\over{\partial N_e}}
\right)
\label{A.0.12}
\end{equation}

\begin{equation}
{\partial^2 F_2 \over{\partial V \partial N_q} }
= - {\kbt\over Z_q} 
{\partial Z_q\over{\partial V}}
\label{A.0.13}
\end{equation}

%\begin{equation}
%{\partial Q_{is}\over{\partial N_e}} = Q_{is}^{'} 
%{\partial \beta_{is}\over{\partial N_e}} = - {2\beta_{is}\over{3 N_e}}
%Q_{is}^{'}
%\label{A.9.1}
%\end{equation}

\begin{equation}
{\partial Z_s\over{\partial N_e}} = 
- {2\over{3 N_e}} \sum_i {\beta_{is} Q_{is}^{'}\over{Q_{is}}}
w_{is}g_{is}e^{{-\Delta E_{is}/{\kbt}} }
\label{A.1.1}
\end{equation}

\begin{equation}
{\partial^2 Z_s\over{\partial N_e^2}} = - 
{\left(\frac{2}{3 N_e}\right)}^2 \sum_i {\beta_{is}^2\over{Q_{is}}}
{\left({Q_{is}^{''} + 
{5\over2}{Q_{is}^{'}\over{\beta_{is}}}}\right)}w_{is}g_{is}
e^{{-\Delta E_{is}/{\kbt}} }
\label{A.1.2}
\end{equation}

\begin{equation}
{\partial Z_\nu\over{\partial N_\mu}} = 
- {4\pi\over{3 V}} 
\sum_i {(r_{i\nu} + r_{1\mu})}^3
w_{i\nu}g_{i\nu}e^{{-\Delta E_{i\nu}/{\kbt}} }
\label{A.1.3}
\end{equation}

\begin{equation}
{\partial^2 Z_\nu\over{\partial N_\mu \partial N_\lambda}} =
{\left({4\pi\over{3 V}}\right)}^2 
\sum_i {(r_{i\nu} + r_{1\mu})}^3 {(r_{i\nu} + r_{1\lambda})}^3
w_{i\nu}g_{i\nu}e^{{-\Delta E_{i\nu}/{\kbt}} }
\label{A.1.4}
\end{equation}

\begin{equation}
{\partial^2 Z_\nu\over{\partial N_\mu \partial N_e}} =
{8\pi\over{9 N_{e} V}}
\sum_i {(r_{i\nu} + r_{1\mu})}^3 
{\left({ 
\beta_{i\nu} Q_{i\nu}^{'} \over{Q_{i\nu}}
}\right)} 
w_{s\nu}g_{i\nu}e^{{-\Delta E_{i\nu}/{\kbt}} }
\label{A.1.5}
\end{equation}

\begin{equation}
{\partial Z_s\over{\partial T}} = 
{1\over{\kbt^2}} \sum_i {\Delta E_{is}}
w_{is}g_{is}e^{{-\Delta E_{is}/{\kbt}} }
\label{A.1.6}
\end{equation}

\begin{equation}
{\partial^2 Z_s \over{\partial T^2} } =
{\left({1\over{\kbt^2}}\right)}^2 
\sum_i {\Delta E_{is}}
{\left({ \Delta E_{is} - 2 \kbt }\right)}
w_{is}g_{is}e^{{-\Delta E_{is}/{\kbt}} }
\label{A.1.7}
\end{equation}

\begin{equation}
{\partial^2 Z_s \over{\partial T \partial N_e} } =
- {2\over{3 \kbt^2 N_{e}}}
\sum_i {\Delta E_{is}}
{\left({ 
\beta_{is} Q_{is}^{'} \over{Q_{is}}
}\right)}
w_{is}g_{is}e^{{-\Delta E_{is}/{\kbt}} }
\label{A.1.8}
\end{equation}

\begin{equation}
{\partial^2 Z_s \over{\partial T \partial N_\mu} } =
- {4 \pi \over{3 \kbt^2 V}}
\sum_i {\Delta E_{is}}
{(r_{is} + r_{1\mu})}^3 
w_{is}g_{is}e^{{-\Delta E_{is}/{\kbt}} }
\label{A.1.9}
\end{equation}

\begin{equation}
{\partial Z_s \over{\partial V} } =
{1\over{V}}
\sum_i 
\left[
{2 \beta_{is} Q_{is}^{'}\over{3 Q_{is}}} +
\frac{4 \pi}{3V} \sum_{\nu} N_{\nu}(r_{is}+r_{1\nu})^{3}
\right]
w_{is}g_{is}e^{-\Delta E_{is}/{\kbt}} 
\label{A.1.10}
\end{equation}

\begin{equation}
{\partial^2 Z_s \over{\partial T \partial V} } =
{1\over{\kbt^{2} V}} \sum_i \Delta E_{is}
{\left[{ {2 \beta_{is} Q_{is}^{'}\over{3 Q_{is}}} +
\frac{4 \pi}{3V} \sum_{\nu} N_{\nu}(r_{is}+r_{1\nu})^{3} }\right]}
w_{is}g_{is}e^{{-\Delta E_{is}/{\kbt}} }
\label{A.1.11}
\end{equation}

\begin{flushleft}
$${\partial^2 Z_s \over{\partial V \partial N_\mu} } =
\frac{4\pi}{3 V^{2}} \sum_i{(r_{is} + r_{1\mu})}^3
\hspace{70mm}$$
\end{flushleft}

\vspace{-10mm}
\begin{equation}
\;\;\;\;\;\;\;\;\;\;\;\;\;\;\;\;\;\;\;\;\;\;\;\;\;\;\;\;\;\;\;\;\;\;\;\;
{\left\lbrace{
-{2\over 3} { \beta_{is} Q_{is}^{'}\over{ Q_{is}}} +
{\left[{ 
1 - {4\pi\over{3 V}} \sum_{\nu} N_\nu
{(r_{is} + r_{1\nu})}^3 
}\right]}
}\right\rbrace}
w_{is}g_{is}e^{{-\Delta E_{is}/{\kbt}} }
\label{A.1.12}
\end{equation}

\begin{flushleft}
$${\partial^2 Z_s \over{\partial V \partial N_e} } =
- \frac{2}{3 N_{e} V}\sum_i
\left\{ {\frac{4\pi}{3V}{\beta_{is} Q_{is}^{'}\over{Q_{is}}}
\left[\sum_{\nu} N_{\nu}(r_{is}+r_{1\nu})^{3}\right]}
\right.\hspace{60mm}$$
\end{flushleft}

\vspace{-10mm}
\begin{equation}
\;\;\;\;\;\;\;\;\;\;\;\;\;\;\;\;\;\;\;\;\;\;\;\;\;\;\;\;\;\;\;\;\;\;\;\;
\;\;\;
\left.
+ \frac{2}{3} \sum_i {\beta_{is} \over{ Q_{is}}} (Q_{is}^{'}
+ \beta_{is} Q_{is}^{''})\right\}
w_{is}g_{is}e^{{-\Delta E_{is}/{\kbt}} }
\label{A.1.13} 
\end{equation}

\begin{flushleft}
$${\partial^2 Z_s \over{\partial V^2 } } =
{1\over V^2} \sum_i\left\{
\left[\frac{4 \pi}{3V}\sum_{\nu} N_{\nu}(r_{is}+r_{1\nu})^{3}\right]
\left[ {4\over 3} {\beta_{is} Q_{is}^{'}\over{Q_{is}}}
+ \left(\frac{4\pi}{3V}\sum_{\nu} N_{\nu}(r_{is}+r_{1\nu})^{3}-2\right) \right]
\right.
\hspace{10mm} $$
\end{flushleft}

\vspace{-10mm}
\begin{equation}
\;\;\;\;\;\;\;\;\;\;\;\;\;\;\;\;\;\;\;\;\;\;\;\;\;\;\;\;\;\;\;\;\;\;\;\;
\;\;\;\;\;\;\;\;\;\;\;\;\;\;\;\;\;\;\;\;\;\;
\left.
+ {4\over 9}{\beta_{is}\over{Q_{is}}}
(\beta_{is}Q_{is}^{''}-{1\over 2} Q_{is}^{'})\right\}
w_{is}g_{is}e^{-\Delta E_{is}/{\kbt}}
\label{A.1.14}
\end{equation}

\begin{table}
\caption{
Properties of the solar models of Fig.~\ref{fig-hel3}. See text.
}
\begin{tabular}{llllll}
\hline
Model & \hbox{EOS} & \hbox{Radius} & \hbox{Convective}& $Y_s$& $r_{\rm
      cz}/R_\odot$ \\
      & & \hbox{Mm} & \hbox{Flux}& & \\
\hline
M1 & MHD & 695.78 & CM & 0.2472 & 0.7145 \\
M2 & MHD & 695.99 & CM & 0.2472 & 0.7146 \\
M3 & MHD & 695.51 & CM & 0.2472 & 0.7145 \\
M4 & MHD & 695.78 & MLT & 0.2472 & 0.7146 \\
M5 & OPAL& 695.78 & CM & 0.2465 & 0.7134 \\
M6 & OPAL& 695.99 & CM & 0.2465 & 0.7135 \\
M7 & OPAL& 695.51 & CM & 0.2466 & 0.7133 \\
M8 & OPAL& 695.78 & MLT & 0.2465 & 0.7135 \\
\hline
\end{tabular}
\end{table}

\clearpage

\clearpage

\figcaption{{\it Left Panel:} 
absolute values of $\gamma_1$ for solar temperatures and densities
of a hydrogen-only plasma. Linestyles: MHD -- asterisks, ${\rm MHD_{GS}}$ --
dashed lines, ${\rm MHD_{PL}}$ -- dotted-dashed lines, ${\rm MHD_{PL,GS}}$ --
dotted lines, and OPAL -- solid lines. 
See text for the definitions of the different MHD versions.
{\it Right Panel:} relative differences with
respect to ${\rm MHD_{GS}}$, in the sense $(\gamma_1 - \gamma_1[{\rm
MHD_{GS}}])/\gamma_1[{\rm MHD_{GS}}])$, using the same line styles as in 
the left panel.
The horizontal solid zero line, representing ${\rm MHD_{GS}}$, is also shown.
\label{fig-hel2}}
\vspace{4cm}
\plottwo{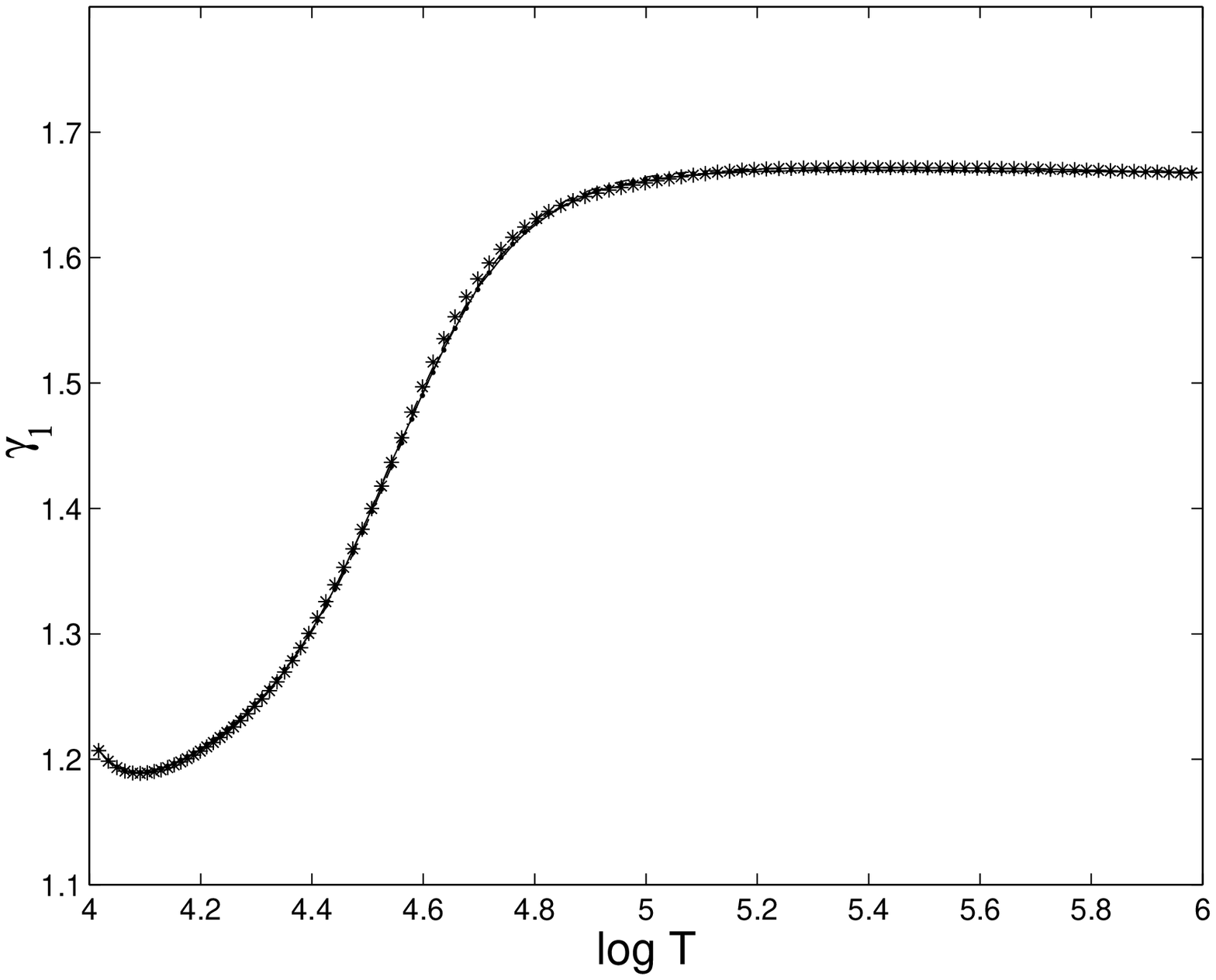}{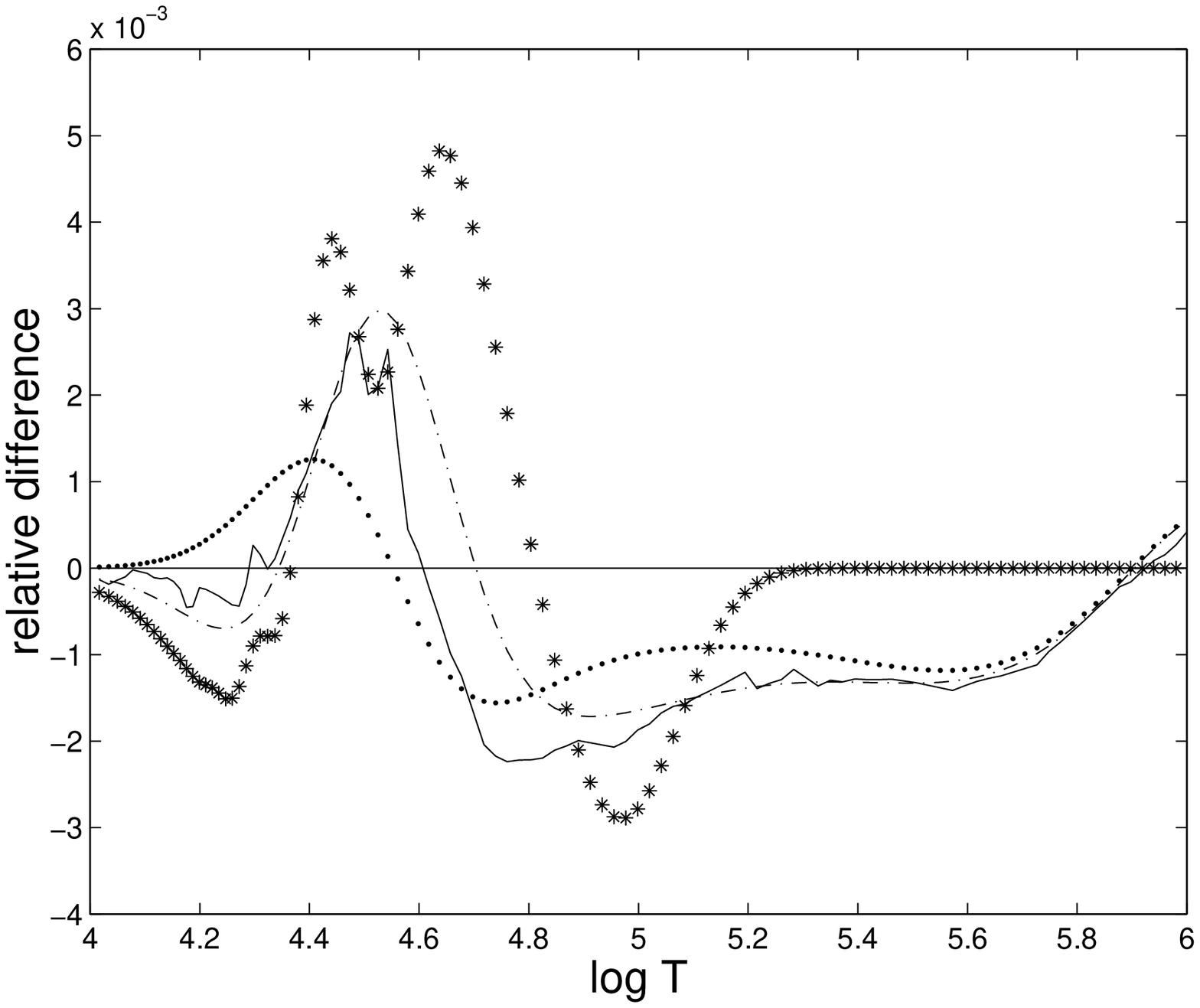}

\clearpage

\clearpage

\clearpage

\clearpage

\clearpage

\clearpage

\clearpage

\clearpage

\clearpage

\clearpage

\clearpage

\clearpage

\clearpage

\clearpage

\clearpage

\clearpage

\end{document}